%% file: breathing_monitoring.tex
\documentclass[journal, 10pt, twocolumn, a4paper]{IEEEtran}
\IEEEoverridecommandlockouts    % to create the author's affliation portion
\usepackage{stmaryrd}
\usepackage{amsfonts}
\usepackage{amssymb}
\usepackage{amsmath}
\usepackage{amsthm}
\usepackage{mathrsfs}
\usepackage{verbatim}

	\usepackage[pdftex]{graphicx}
	\graphicspath{{./}}
	\DeclareGraphicsExtensions{.pdf,.jpeg,.png}
	\usepackage[pdftex, bookmarks = true, colorlinks=false]{hyperref}
\RequirePackage{lineno}

\usepackage{color}
\usepackage[usenames]{xcolor}

\usepackage{arydshln,leftidx,mathtools}
\usepackage{booktabs}
\usepackage{subfig}

\DeclareMathOperator*{\argmax}{arg\,max}
\DeclareMathOperator*{\diag}{diag}

\begin{document}

\setlength{\dashlinegap}{1.0pt}
%\setpagewiselinenumbers
%\modulolinenumbers[3]

%\linenumbers
%\linenumbersep 3pt \relax

\input{main.tex}



% http://www.ctan.org/tex-archive/macros/latex/contrib/supported/IEEEtran/testflow/bibtex
%\bibliographystyle{IEEEtran.bst}

\bibliographystyle{IEEEtran}
\end{document}

%% file: main.tex
% paper title: Must keep \ \\ \LARGE\bf in it to leave enough margin.
\title{\ \\ \LARGE\bf Catch a Breath: Non-invasive Respiration Rate Monitoring \\
via Wireless Communication
\thanks{Ossi Kaltiokallio, H\"{u}seyin~Yi\u{g}itler, and Riku~J\"{a}ntti are with the Department of Communications and Networking, Aalto University, School of Electrical Engineering, Espoo, Finland (email:\{name.surname\}@aalto.fi). Neal Patwari is with the Department of Electrical and Computer Engineering, University of Utah, Salt Lake City, Utah, USA (email:npatwari@ece.utah.edu).}}

%This work was supported by the Finnish Funding Agency for Technology and Innovation under project WISM II }}

\author{Ossi Kaltiokallio, H\"{u}seyin~Yi\u{g}itler, Riku~J\"{a}ntti and Neal Patwari}

% avoiding spaces at the end of the author lines is not a problem with
% conference papers because we don't use \thanks or \IEEEmembership
%\specialpapernotice{(Invited Paper)}

\maketitle

\begin{abstract}
Radio signals are sensitive to changes in the environment, which for example is reflected on the received signal strength (RSS) measurements of low-cost wireless devices. This information has been used effectively in the past years e.g. in device-free localization and tracking. Recent literature has also shown that the fading information of the wireless channel can be exploited to estimate the breathing rate of a person in a non-invasive manner; a research topic we address in this paper. To the best of our knowledge, we demonstrate for the first time that the respiration rate of a person can be accurately estimated using only a single IEEE 802.15.4 compliant TX-RX pair. We exploit channel diversity, low-jitter periodic communication, and oversampling to enhance the breathing estimates, and make use of a decimation filter to decrease the computational requirements of breathing estimation. In addition, we develop a hidden Markov model (HMM) to identify the time instances when breathing estimation is not possible, i.e., during times when other motion than breathing occurs. We experimentally validate the accuracy of the system and the results suggest that the respiration rate can be estimated with a mean error of 0.03 breaths per minute, the lowest breathing rate error reported to date using IEEE 802.15.4 compliant transceivers. We also demonstrate that the breathing of two people can be monitored simultaneously, a result not reported in earlier literature.
\end{abstract}

\section{Introduction} \label{S:introduction}
\PARstart{T}{he success} of wireless communication systems together with the recent advances in different technologies have enabled the development of wireless sensor networks (WSNs) \cite{Akyildiz2002}. These networks are composed of low-cost transceivers and currently WSNs are used and tested in different application areas such as wireless control \cite{kaltiokallio2011}, structural health monitoring \cite{Ceriotti2009}, and health care \cite{Ko2010b}. In addition, WSNs are finding their way into a new type of sensing where the wireless medium itself is probed using the communications of a dense network deployment. Such networks are referred to as RF sensor networks \cite{patwari2010} since the radio of the low-cost transceivers is used as the sensor. These networks do not require people to co-operate with the system, allowing one to gain situational awareness of the environment non-invasively. Consequently, RF sensor networks are rendering new sensing possibilities such as device-free localization (DFL) \cite{Wilson2010}, and non-invasive breathing monitoring \cite{Patwari2011}.  

Wireless networks are ubiquitous nowadays. Wherever we are, we interact with radio signals by shadowing, reflecting, diffracting and scattering multipath components as they propagate from the transmitter to receiver \cite[pp. 47-67]{Molisch2010}. As a consequence, the channel properties change due to temporal fading \cite{hashemi93}, providing information about location of the interacting objects and about the rate at which the wireless channel is altered. To quantify these changes in the propagation medium, one could for example measure the channel impulse response (CIR) \cite{patwari2010}.

The CIR allows one to measure the amplitude, time delay, and phase of the individual multipath components, but requires the use of sophisticated devices. In the context of situational awareness, the time delay is the most informative. For example, in the simplest scenario when there exists one multipath component in addition to the line-of-sight (LoS) path, the excess delay of the reflected component specifies that an object is located on an ellipse with the TX and RX located at the foci \cite{chang04}. Furthermore, the difference between the excess delays of consecutive receptions determines the rate at which the wireless channel is changing. 

Devices capable of measuring the CIR can be prohibitively expensive, especially when compared to low-cost narrowband transceivers. As a drawback, these low-complexity narrowband devices are only capable of measuring the received signal strength (RSS) which is a magnitude-only measurement. Nevertheless, also the RSS provides information about the surrounding environment. First, when a dominating LoS component is blocked, RSS tends to decrease, indicating that a person is located in between the TX-RX pair \cite{Wilson2010}. Second, variance of the RSS indicates changes in multipath fading \cite{Wilson2011} and therefore, about the location of people and the rate at which they are interacting with the propagation medium.

Despite the fact that narrowband transceivers are not as informative as devices capable of measuring the CIR, one can leverage low-cost of the devices and deploy them in numbers to gain situational awareness. For example, temporal fading information from a dense RF sensor network can be exploited to perform DFL \cite{Wilson2010,kaltiokallio2012}. Moreover, recent literature has demonstrated that an RF sensor network can be used to monitor even small changes in the environment such as breathing of a person \cite{Patwari2011,Patwari2013}; a research topic we address in this paper. 

Inhaling and exhaling of a breathing person causes very small variations in the propagation channel, which is reflected in the RSS measurements as small changes. A typical low-cost narrowband receiver's RSS measurement circuitry is composed of low-quality analog electronics and a quantizer. Thus, the induced noise can hide the breathing signal. In addition, breathing is typically observable in the RSS measurements of those links which are also the most sensitive to other variations in the environment, e.g. to movement of people. These other variations in the wireless channel cause undesired temporal fading which leaks to the RSS measurements and hides the breathing-induced RSS changes. We refer to such variations in the RSS as \textit{motion interference}, since they degrade the performance of breathing monitoring.

In this paper, we address the above two challenges by designing a breathing monitoring system which makes the following contributions:
\begin{itemize}
  \item Inexpensive IEEE 802.15.4 compliant transceivers are used as opposed to more expensive and complex ultra-wideband (UWB) radios \cite{rivera2006} or Doppler radars \cite{Li2010} which are also RF-based.
  \item It is empirically demonstrated that breathing can be accurately estimated using only one TX-RX pair, contrary to other related works which use numerous of sensors to fulfill the task \cite{Patwari2011,Patwari2013}.
  \item Channel diversity, low-jitter periodic communication, and oversampling are exploited to enhance the breathing estimates and decimation is used to decrease the computational demands of breathing estimation.
  \item A hidden Markov model (HMM) is developed to identify motion interference, i.e., time instances when breathing estimation is not possible.
  \item Despite the simplicity of the experimental setup, we report the lowest breathing rate errors to date compared to other works \cite{Patwari2011,Patwari2013} which also use IEEE 802.15.4 compliant transceivers.
  \item We also demonstrate that the breathing of two people can be monitored simultaneously, a result not reported in earlier literature.
\end{itemize}

The rest of the paper is organized as follows. In the remainder of this section we discuss the related work. Section \ref{S:methodology} introduces the overview and requirements of breathing monitoring, the breathing-induceded RSS model, and individual components of the proposed breathing monitoring system. Section \ref{S:experiments} describes the experimental setup that is used to validate the performance of the proposed system and Section \ref{S:results} presents the results of the experiments. Conclusion are drawn in Section \ref{sec:conlcusions}.

\subsection{Related Work}\label{sec:related_work}

Wireless technologies are finding their way into non-invasive vital sign monitoring, and common approaches include using a Doppler radar \cite{Chen1986} or UWB transceivers \cite{Salmi2011} to monitor the respiration rate of a breathing person. Moreover, UWB \cite{rivera2006} as well as Doppler radar \cite{Li2010} have demonstrated the potential to simultaneously monitor the heart rate. These works have shown that the vital signs can be monitored accurately in a non-invasive manner. As a drawback, they rely on sophisticated and expensive hardware making them impractical in many applications. As an example, the Kai Medical ``Continuous`` respiratory rate monitor \cite{Kai}, although not yet FDA approved, is based on Doppler radar and said to be priced at $\$2000$. 
 
As an inexpensive alternative, one can use off-the-shelf narrowband transceivers for estimating the respiration rate of a person. RSS-based breathing monitoring was introduced in \cite{Patwari2011} and experimentally validated on a ``patient`` in a hospital bed. Moreover, an RF sensor network capable of monitoring and locating a breathing inhabitant in a residential apartment was demonstrated in \cite{Patwari2013}. These systems make measurements between $N$ transceivers and one can increase the capability of the network by using more sensors, since the number of measurements increases by $\mathcal{O}(N^2)$. 

Non-invasive vital sign monitoring can create new opportunities not only for improving patient monitoring in hospitals but also in home healthcare e.g. to diagnose and monitor obstructive sleep apnea and sudden infant death syndrome \cite{Droitcour2004}. Other opportunities include: enhancing the life quality of elderly in ambient assisted living applications, to add context-awareness in smart homes, and in search and rescue for earthquake and fire victims.

\section{Methodology} \label{S:methodology}
\subsection{Breathing Monitoring System}

\begin{figure*}[!th]
\begin{centering}
\begin{tabular}{cc}
\mbox
{
\subfloat[]{\includegraphics[width=\columnwidth]{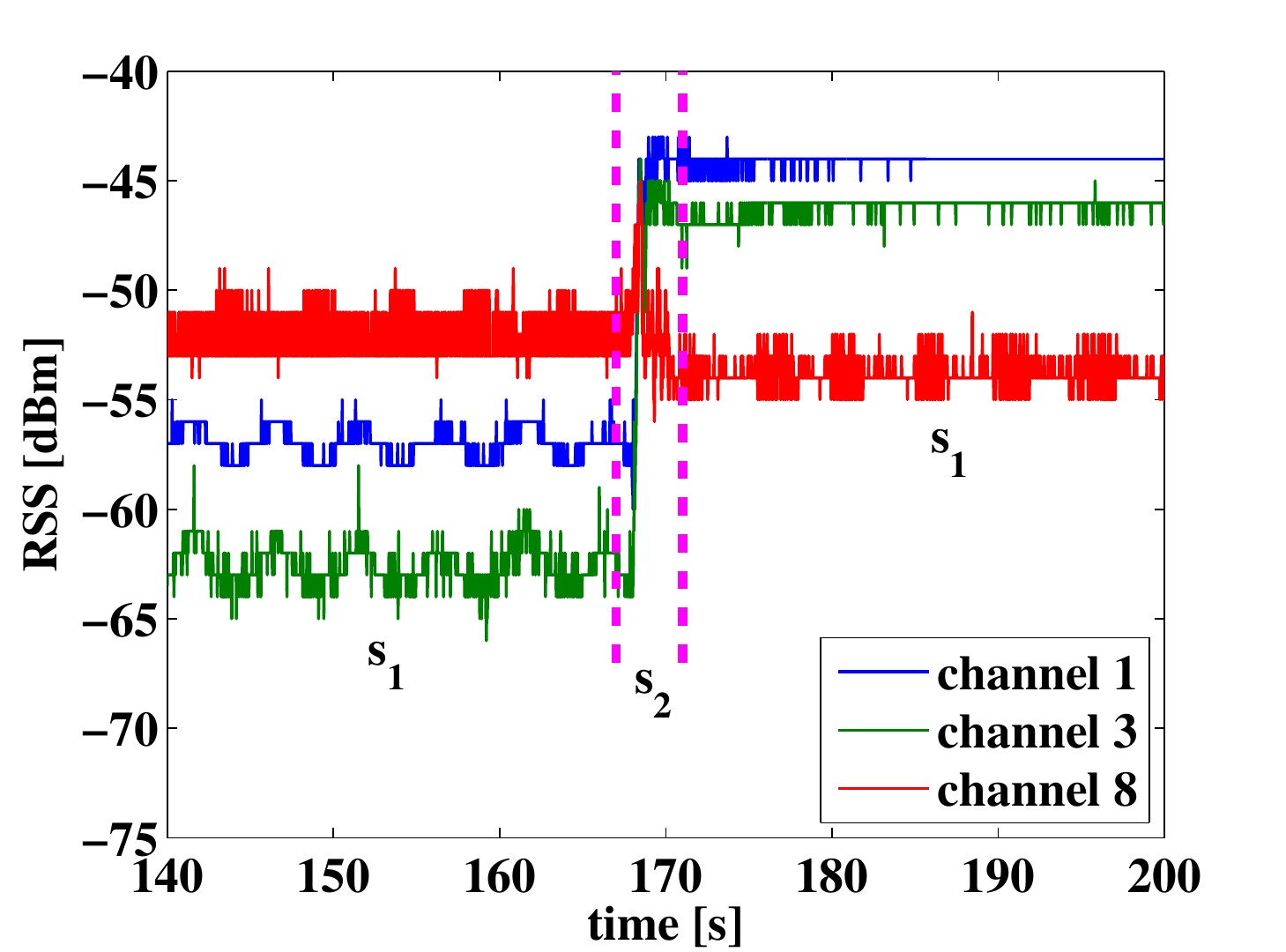}}
\subfloat[]{\includegraphics[width=\columnwidth]{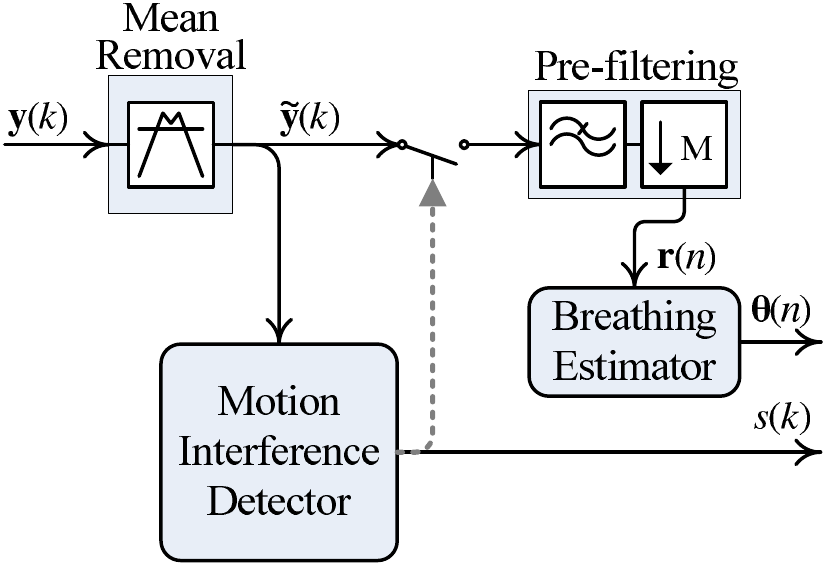}}
}
\end{tabular}
\caption{In (a), RSS measurements on three different channels and in (b), overview and components of the proposed breathing monitoring system.} 
\label{fig:system_overview}
\end{centering}
\end{figure*}

The respiration rate of a person can be monitored during time intervals of no motion interference and if the resolution of the RSS measurements is high enough. As an example, variations of the RSS measurements of a single IEEE 802.15.4 compliant TX-RX pair communicating on three different frequency channels at the $2.4 \text{ GHz}$ ISM band are shown in Fig.~\ref{fig:system_overview} (a). Clearly, some of the channels contain a periodic component when the process is stationary. On the contrary, the breathing-induced signal is not resolvable from the RSS measurements during motion interference ($t\approx170s$). It is also to be noted that after motion interference, channels exhibiting the periodic breathing signal have changed. Thus, breathing estimation is sensitive to the surrounding environment and it is expected that the number of wireless links and/or frequency channels capturing the breathing-induced signal are sparse.

\begin{figure}
\begin{centering}
\begin{tabular}{c}
\includegraphics[width=\columnwidth*2/3]{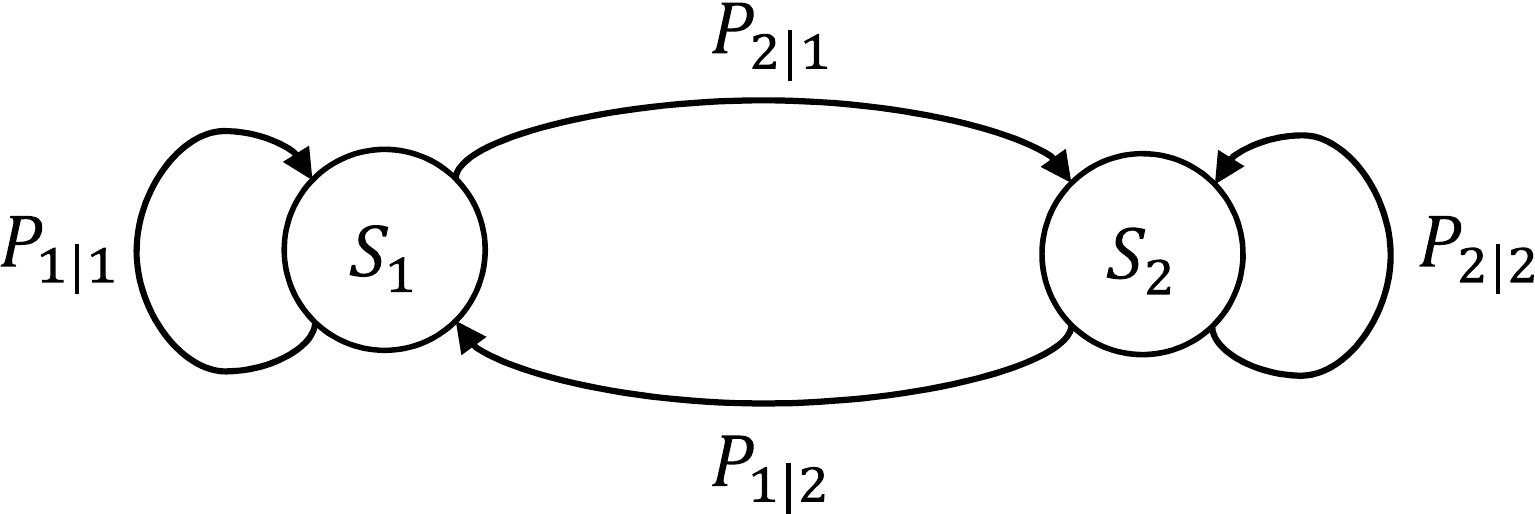}
\end{tabular}
\caption{The two states of the proposed measurement system.} 
\label{fig:states}
\end{centering}
\end{figure}

Based on the observations, three mutually exclusive states for the breathing monitoring system are identified: \textit{no breathing person present} ($S_0$), \textit{no motion interference} ($S_1$), and \textit{motion interference} ($S_2$). In this work however, we assume that a breathing person is always present and therefore, $S_0$ is not considered in remainder of the paper. The two states and the associated state transitions are shown in Fig.~\ref{fig:states}. It is assumed that the current state of the system depends only on the previous state so that the measurement setup can be represented by a two-state Markov chain.

In order to accurately estimate the respiration rate of a person when the process is stationary, a breathing monitoring system depicted in Fig.~\ref{fig:system_overview} (b) is proposed. First, \textit{Mean Removal} is mandatory because of the used spectral estimation technique. Second, a \textit{Motion Interference Detector} monitors the mean removed RSS measurements, denoted by $\boldsymbol{\tilde{y}}(k)$, and the state of the system to enable breathing estimation only when the system is in state $S_1$. Third, $\boldsymbol{\tilde{y}}(k)$, is pre-filtered to enhance the quality of low-resolution measurements and down-sampled by a decimation factor $M$ to decrease the computational requirements of the \textit{Breathing Estimator}. The down-sampled signal, $\boldsymbol{r}(n)$, is used to estimate the parameters, denoted by $\boldsymbol{\theta}$, of the breathing-induced signal.

% ---------------------------------------------
% 		MEASUREMENT MODEL
% ---------------------------------------------
\subsection{Measurement Model}

In the following, we present a breathing induced RSS-model. The RSS measurement in dBm at time $k$ on channel $c$ containing an additional signal can be written as
\begin{equation} \label{eq:rss}
    y_c(k) =  g_c(k) + \epsilon_c(k),
\end{equation}
where $g_c(k)$ is an unknown signal, and $\epsilon_c(k)$ is wide sense stationary (WSS) noise with mean $\mu_c$ and variance $\sigma^2$. When a breathing person is present, it can be assumed that $g_c(k)$ is sinusoidal \cite{Patwari2011},
\begin{equation} \label{eq:rss_breathing}
    g_c(k) = A_c \cos(2 \pi f T_s k+\phi_c),
\end{equation}
where $A_c$, $\phi_c$, and $f$ are the amplitude, phase and frequency in respective order and $T_s$ is the sampling interval. Considering the limitation of resolving a signal from a sequence of measurements, which is dictated by the Nyquist rate, it is not difficult to communicate (sample) at such frequencies that enable breathing estimation since $f \ \text{is near} \ 0.23$ Hz for adults \cite{sebel85}, whereas for newborns $f \ \text{is near} \ 0.62$ Hz \cite{murray86}.

The considered transceivers enable communication over multiples of frequency channels. Thus, the RSS measurement in Eq.~\eqref{eq:rss} can be extended to a measurement vector
\begin{equation}
	\boldsymbol{y}(k) = \left[ \begin{array}{l l c l} y_1(k)&y_2(k)&\cdots&y_C(k) \end{array}\right]^T,
\end{equation}
where $C$ is the number of used channels. Since the periodic component is generated by the breathing person, we assume that the frequency of the sinusoidal signal on the different channels is the same, whereas the amplitude and phase are expected to be channel dependent. Furthermore, the measurement noise contaminating each channel is assumed to be independent of the others. Consequently, the measurement model of the studied system is given by
\begin{equation}  \label{eq:measurementmodel}
\begin{aligned}
	\boldsymbol{y}(k) &= \boldsymbol{g}(k) + \boldsymbol{\epsilon}(k) \\
	\boldsymbol{g}(k) &= \left[\begin{array}{llcl} g_1(k)&g_2(k)&\cdots&g_C(k) \end{array}\right]^T \\
	\boldsymbol{\epsilon}(k) &= \left[\begin{array}{llcl} \epsilon_1(k)&\epsilon_2(k)&\cdots&\epsilon_C(k) \end{array}\right]^T \\
	\boldsymbol{\mu} &= \left[\begin{array}{llcl} \mu_1&\mu_2&\cdots&\mu_C \end{array}\right]^T. % \\
	%\boldsymbol{\Sigma} &= \sigma^2 \boldsymbol{I},
\end{aligned}
\end{equation}
%where $\boldsymbol{\Sigma}$ is the covariance matrix of the joint distribution. 
Breathing monitoring aims at estimating $\boldsymbol{g}(k)$ using $\boldsymbol{y}(k)$.

\begin{figure*}[!th]
\begin{centering}
\begin{tabular}{ccc}
\mbox
{
\subfloat[]{\includegraphics[width=\columnwidth]{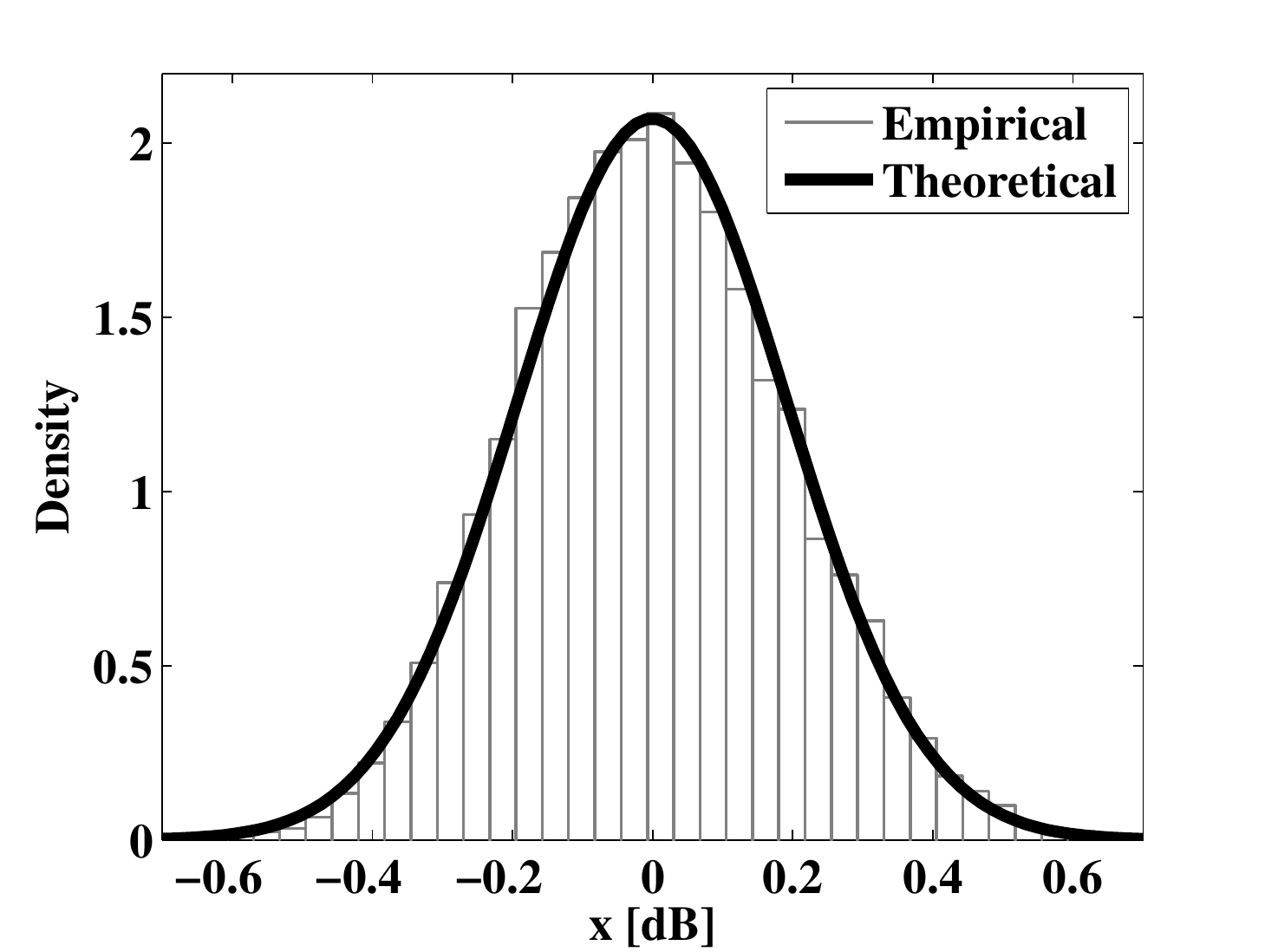}}
\subfloat[]{\includegraphics[width=\columnwidth]{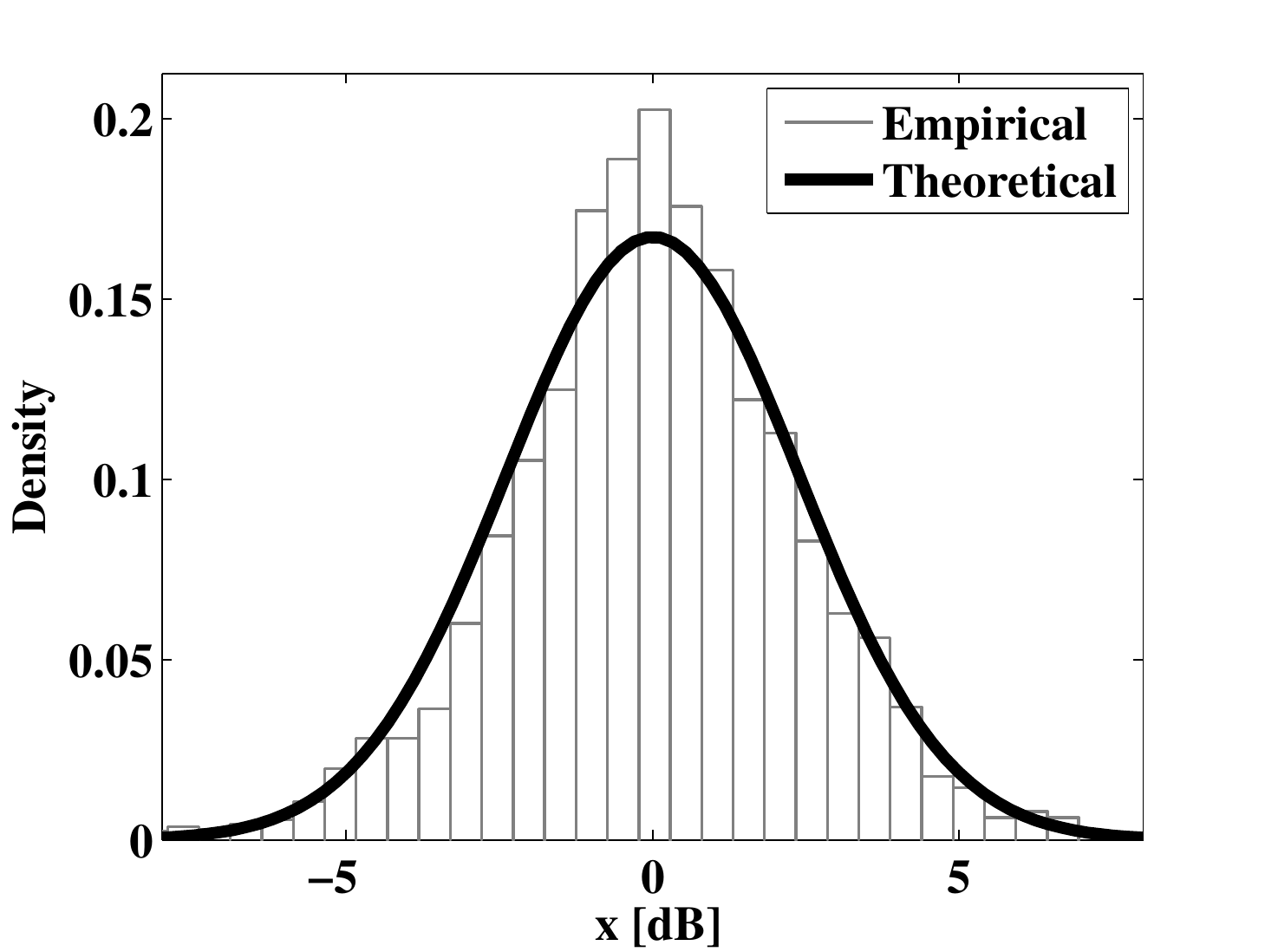}}
}
\end{tabular}
\caption{In (a), the empirical and theoretical conditional densities of the observations when no motion interference is present. Correspondingly, the conditional densities in the presence of motion interference shown in (b).} 
\label{fig:distributions}
\end{centering}
\end{figure*}

% ---------------------------------------------
% 		MEAN REMOVAL
% ---------------------------------------------
\subsection{Mean Removal}
The mean removal subsystem provides zero-mean RSS measurements for the other components of the breathing monitoring system. In \cite{Patwari2013}, the use of a windowed average instead of a 7$^{\text{th}}$ order Chebychev high-pass filter \cite{Patwari2011} was found superior for the purpose of breathing estimation, i.e.,
\begin{equation} \label{eq:average}
     \boldsymbol{\mu} = \frac{1}{L}\sum_{i=0}^{L-1} \boldsymbol{y}(k-i),
\end{equation}
where $L$ is the length of the window. Thus, the output of the mean removal subsystem is $\boldsymbol{\tilde{y}}(k) = \boldsymbol{y}(k) - \boldsymbol{\mu}$.

Considering the two possible states of the system, the window length $L$ must be determined according to parameters of the measurement system. When in state $S_1$, values of $L$ that are too small can suppress the spectral components of $\boldsymbol{g}(k)$. Selecting $L$ too large can allow frequencies much lower than the minimum breathing rate to remain in $\boldsymbol{\tilde{y}}$. When in state $S_2$, $L$ must compensate for the rapid changes in $\boldsymbol{y}(k)$. Therefore, $L$ should be determined considering the sampling frequency $f_s$, and the lowest and highest frequencies the system is designed to detect, i.e., $f_{min}$ and $f_{max}$. In this paper we set $L = f_{s}/f_{max}$.

% ---------------------------------------------
% 		MOTION INTERFERENCE
% ---------------------------------------------
\subsection{Motion Interference Detector}\label{sec:motion_interference}

It has already been shown that the RSS experiences time intervals of considerable fading caused by the movements of people, whereas most of the time, the RSS remains nearly constant \cite{bultitude1987}. This fading / non-fading time varying process can be modeled as a two-state Markov chain \cite{roberts1995}. However, the states of our system are not directly observable and therefore, we represent the system using a hidden Markov model (HMM). In order to make the decision of enabling or disabling breathing estimation, the state of the system is estimated through an observable measure, the RSS, which is a probabilistic function of the unobservable state. 

The HMM can be used to calculate the probability of an observation $x(k)$ given the state transition probabilities $P$, the conditional densities of the observations $p_{x \vert s}$, and the initial state probability $\pi$ using the \emph{forward procedure} \cite[pp. 109-114]{Therrien1992}
\begin{equation} \label{eq:forward_procedure}
     p_x(x(k) \vert P, p_{x \vert s}, \pi) = \sum_{i=1}^{Q}\alpha_i(k),
\end{equation}
where $\alpha$ is the forward variable and $Q$ is the number of states ($Q=2$). The forward variable at time instant $k$ for state $s_i$ can be calculated recursively
\begin{equation} \label{eq:forward_variable}
     \alpha_i(k) = \left[\sum_{j=1}^{Q}\alpha_j(k-1) \cdot P_{i \lvert j}\right]p_{x \vert s}(x(k)\lvert s_i),
\end{equation}
where $\alpha_i(1) = \pi_{s_i} \cdot p_{x \vert s}(x(1)\lvert s_i)$. The output of the motion interference detector is the state that has the highest probability at time $k$ and it is used to enable/disable breathing estimation.

% ---------------------------------------------
% 	     CONDITIONAL DENSITIES
% ---------------------------------------------
\subsection{Conditional Observation Densities}\label{sec:conditional_densities}

In order to determine the probability of the observation at time instant $k$, it is mandatory to consider the distribution of the observations given the state of the system. In this paper, our observation for the HMM is the average of the mean removed RSS measurements
\begin{equation} \label{eq:observation}
     %x(k)= \frac{1}{C} \cdot \boldsymbol{\tilde{y}} (k)^T  \boldsymbol{1}_C,
     x(k)= \frac{1}{C} \sum_{c=1}^{C} \tilde{y}_c(k).
\end{equation}
%
%
%where $\boldsymbol{1}_C$ is all ones vector in $\mathbb{R}^C$. 
This observation model is based on the following properties of the propagation channel and the used measurement setup. First, the time duration of one communication cycle (a single transmission on each frequency channel) is much lower than the coherence time of the channel. Thus, the propagation medium can be considered stationary over a single communication cycle. Second, the coherence bandwidth is assumed to be approximately flat, i.e., fading among the different channels are highly correlated. Consequently, the observation model in Eq.~\eqref{eq:observation} allows us to use well-known distributions to characterize $x(k)$ in the different states.

In state $S_1$, the RSS measurements are dominated by quantization errors and electronic noise. Therefore, $x(k)$ is the sum of independent and identically distributed random variables since the mean is removed and the different frequency channels use the same receiver. Due to the central limit theorem, the density of $x(k)$ is expected to be Gaussian with zero-mean
\begin{equation} \label{eq:gaussian_distribution}
     p_{x \vert s}\left(x(k) ; 0, \sigma\right) = \frac{1}{\sigma \sqrt{2\pi}}\exp \left(-\frac{x(k)^2}{2\sigma^2}\right).
\end{equation}

The Rayleigh, log-normal, Nakagami and Ricean distributions are typically used in describing multipath fading, each having their own theoretical justification \cite{hashemi93}. For example, in obstructed environments the transmitted signal typically experiences several reflections resulting that the observed fading can be characterized as a multiplicative process $-$ giving rise to the log-normal distribution. The RSS measurements during motion interference are expected to follow the distribution describing multipath fading, since fading dominates the other noise sources. We use the log-normal distribution to characterize multipath fading, i.e., the observations when in state $S_2$. Thus, in logarithmic scale, $x(k)$ has the Gaussian density given in Eq.~\eqref{eq:gaussian_distribution}.
%%
%%
%In the presence of motion interference, the RSS measurements are expected to have a Ricean distribution in linear scale \cite{Molisch2010}. However, the densities are expected to be different among the various frequency channels due to e.g. differences in the components of $\boldsymbol{\mu}$. In our measurement system the components of $\boldsymbol{\tilde{y}}(k)$ are zero-mean in logarithmic scale and as a result, $x_k$ is expected to have a density which is characterized by a central chi-squared distribution \cite[pp. 49-53]{Stuber2001}. The central chi-squared distribution is a special case of the gamma distribution for which the probability density function is
%%
%%
%\begin{equation} \label{eq:gamma_distribution}
     %p_{x \vert s}(x ; \beta , \lambda) = \frac{1}{\lambda^\beta}\frac{1}{\Gamma(\beta)}x^{\beta-1}e^{-\frac{x}{\lambda}},
%\end{equation}
%%
%%
%where $\beta$ is the shape parameter, $\lambda$ the scale parameter, and $\Gamma(\cdot)$ the gamma function. This approximation does not only simplify the model, but it also allows us to make the system independent of the measurement setup.

In Section \ref{sec-hmm-fit}, we conduct two experiments to derive the empirical distributions of the observations. The data are fitted to the theoretical density given in Eq.~\eqref{eq:gaussian_distribution}. Statistical analysis is performed to verify goodness of the fits. The empirical and theoretical densities of the observations in the two different states are shown in Fig.~\ref{fig:distributions}.

% ---------------------------------------------
% 		PRE-FILTERING
% ---------------------------------------------
\subsection{Pre-filtering}\label{sec:filtering}
The amplitudes of $\boldsymbol{g}(k)$ are low and therefore, the quantization error of low resolution radio peripherals can hide the sinusoidal breathing signal. One could increase the resolution of the RSS measurements using for example a higher resolution analog-to-digital converter (ADC). Another option is to over-sample and then filter the measurements to increase the resolution \cite[pp. 182-183]{Oppenheim1999}. We adopt the latter option and pre-filter the RSS measurements to increase the resolution of the breathing-induced signal. Further, decimation is exploited to decrease the computational requirements of the breathing estimator. Consequently, the designed pre-filter is a finite impulse response (FIR) decimator. 

Given the range of possible breathing frequencies, the filter is designed to have a passband frequency of $f_{min} =$ 0.1 Hz and a stopband frequency of $f_{max} =$ 1 Hz. The passband ripple of the designed filter is 0.05 dB and has a 40 dB attenuation at frequencies higher than 1 Hz. The filter downsamples the measurements by a decimation factor $M$, thus the sampling interval at the output is $M T_{s}$. We denote the mean removed and filtered RSS measurements as $\boldsymbol{r}(n)$. 

\begin{figure}
\centerline{\includegraphics[width=\columnwidth]{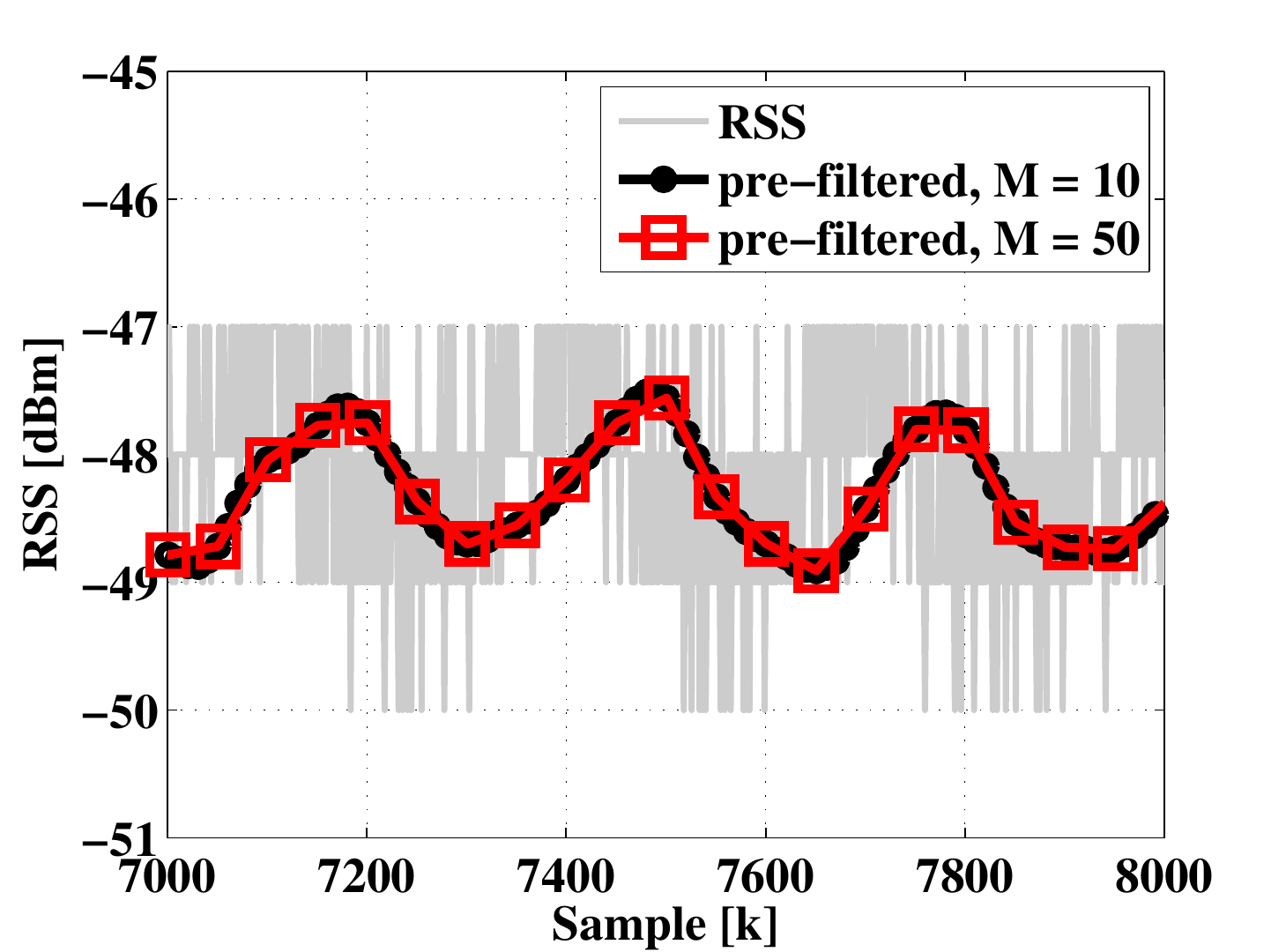}}
\caption{The RSS measurements shown in gray and the pre-filtered signals using a decimation factor of $M=10$ (black) and $M=50$ (red).} \label{fig:decimator}
\end{figure}

In Fig.~\ref{fig:decimator}, the measured RSS and the filtered signal using two different decimation factors is shown. The RSS has a periodic term included but it is not apparent due to quantization errors and noise. When pre-filtering is applied, the breathing-induced signal becomes evident. In section \ref{sec:sampling} we investigate the effect of $M$ to the computational cost and accuracy of the system. 

% ---------------------------------------------
% 	     BREATHING MONITORING
% ---------------------------------------------
\subsection{Breathing Estimator}\label{sec:breathing_monitoring}
Breathing monitoring aims at estimating $\boldsymbol{g}(n)$ using $N$ samples of the pre-filtered and down-sampled signal,
\begin{equation}\label{eq:rss_breathing2}
	\begin{aligned}
	\boldsymbol{r}(n) &= \boldsymbol{g}(n) + \boldsymbol{\eta}(n) \\
	\boldsymbol{\eta}(n) &\sim \mathcal{N}(\boldsymbol{0}, \boldsymbol{\Sigma}) \\
	\boldsymbol{\Sigma} &= \diag\left\{\sigma^2_1,\sigma^2_2, \cdots, \sigma^2_C \right\}
	\end{aligned}
\end{equation}
where $\diag\{\cdot\}$ is the diagonal matrix of the variances. The noise process, $\boldsymbol{\eta}(n)$, is expected to be approximately WSS Gaussian since $M$ independent zero-mean measurements are summed by the decimator. This assumption is validated using the Kolmogorov-Smirnov test \cite{massey1969} using a confidence level of $95\%$. The hypothesis that $\boldsymbol{\eta}(n)$ is Gaussian is accepted for $13$ of the $16$ independent channels with an average p-value of $69\%$ justifying the assumption. 

The parameters of $\boldsymbol{g}(n)$ are estimated using a maximum likelihood estimator (MLE) which is an extension of the standard sinusoid parameter estimator \cite[pp. 193-195]{Kay1993}. If the covariance of the noise process in \eqref{eq:rss_breathing2} is further simplified by assuming that the variances of all the components are the same, the log-likelihood function can be written as
\begin{equation} \label{eq:breathing_estimate}
    J(\boldsymbol{\theta}) = -\frac{1}{2}\sum\limits_{c=1}^{C}\sum\limits_{n=1}^{N}\left(r_c(n) - g_c(n)\right)^2.
\end{equation}
In Eq.~\eqref{eq:breathing_estimate}, $\boldsymbol{\theta}$ denotes the parameters of $\boldsymbol{g}(n)$, i.e.,
\begin{equation} \label{eq:model_parameters}
    \boldsymbol{\theta} = [\boldsymbol{A},\boldsymbol{\Phi},f]^T,
\end{equation}
where $\boldsymbol{A}=[A_1,\ldots, A_C]$ and $\boldsymbol{\Phi}=[\phi_1,\ldots, \phi_C]$ are the amplitude and phase of the different channels, and $f$ is the common frequency. 

The parameters of $\boldsymbol{g}(n)$ can be estimated using spectral estimation techniques as proposed by Patwari \emph{et al.} \cite{Patwari2011}. A good approximation of the MLE of ${f}$ is the frequency where the power spectral density (PSD) has its maximum
\begin{equation} \label{eq:psd}
    \hat{f} = \argmax_{f_{min} \leq f \leq f_{max}} \frac{1}{C}\sum_{c=1}^{C} \Bigl\lvert \sum_{n=1}^{N}r_c(n) e^{-j 2\pi f {T_s} n/M} \Bigr\rvert^2,
\end{equation}
where $r_c(n)$ is the $c^\text{th}$ component of $\boldsymbol{r}(n)$, and $M$ is the decimation factor. After the frequency is estimated, it can be used to compute the estimates of the channel amplitudes and phases using
\begin{eqnarray} 
    \hat{A}_c &=& \frac{2}{N}\Bigl\lvert \sum_{n=1}^{N}r_c(n) e^{-j 2\pi \hat{f} T_s n / M} \Bigr\rvert, \label{eq:amplitude}\\
    \hat{\phi}_c &=& \arctan\frac{-\sum_{n=1}^{N}r_c(n) \sin(2\pi \hat{f} T_s n / M)}{\sum_{n=1}^{N}r_c(n) \cos(2\pi \hat{f} T_s n / M)}.\label{eq:phase}
\end{eqnarray}

\begin{table}
    \caption{Experimental parameters} % title of Table
        \centering % used for centering table
        \begin{tabular}{c c c} % centered columns (4 columns) \\
        \hline\hline\ %inserts double horizontal lines
        Parameter & Value & Description \\
        \hline  % inserts single horizontal line
        $T_s$ & 16  & Sampling rate [ms] \\ % inserting body of the table
	$C$ & 16  & Number of channels \\ % inserting body of the table
	$f_s$ & 62.5  & Sampling frequency [Hz] \\ % inserting body of the table
        $f_{min}$ & 0.1 & Minimum breathing frequency [Hz]  \\ % inserting body of the table
	$f_{max}$ & 1.0 & Maximum breathing frequency [Hz] \\ % inserting body of the table
	$L$ & $f_s/f_{max}$ & Window length of mean removal (1 s)\\ % inserting body of the table
	$M$ & 10 & Decimation factor\\ % inserting body of the table
        \hline %inserts single line
        \end{tabular}
        \label{table:experiment_parameters} % is used to refer this table in the text
\end{table}

\begin{figure}
\centerline{\includegraphics[width=\columnwidth]{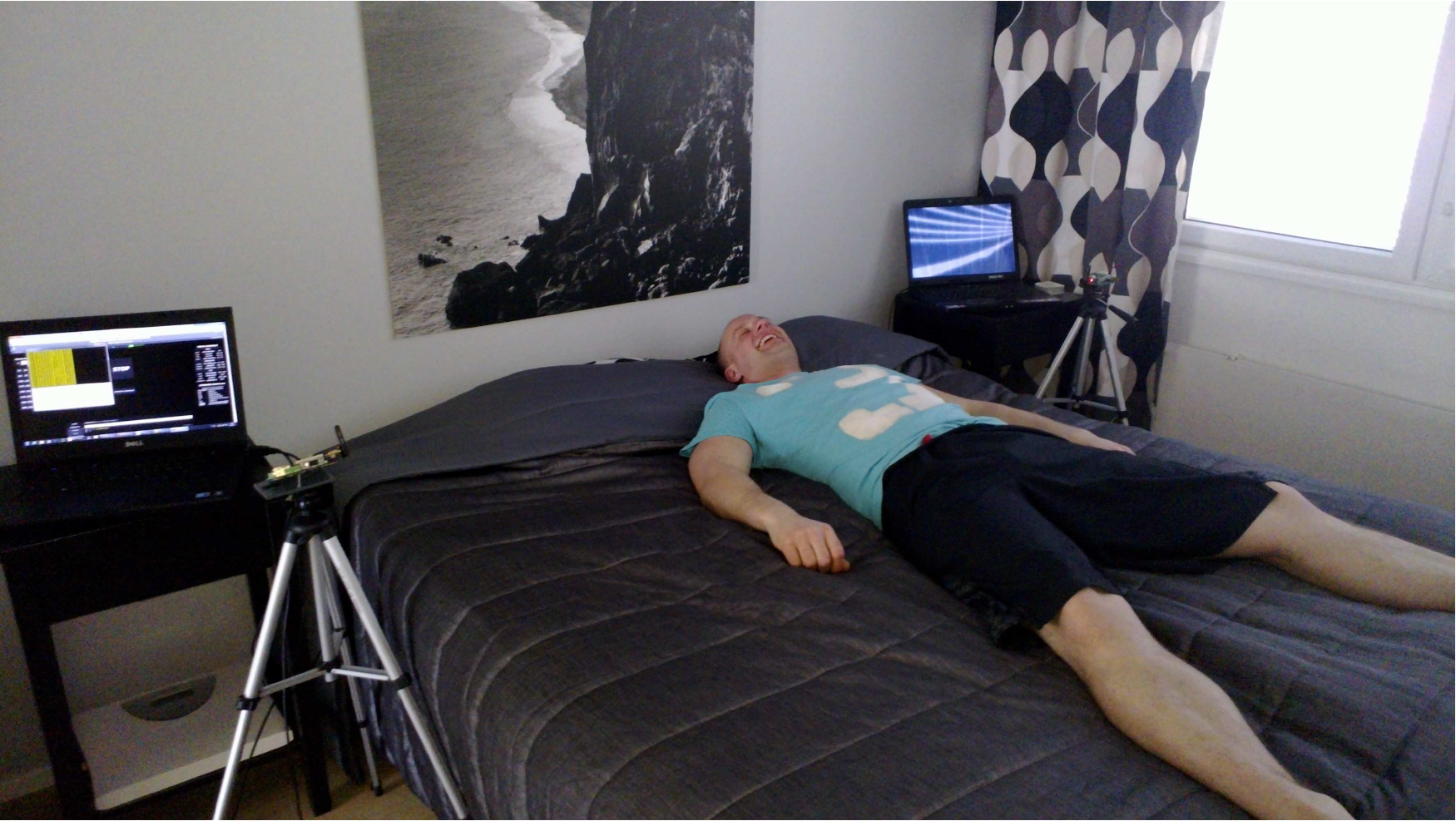}}
\caption{Experiment setup} \label{fig:experiment}
\end{figure}

\section{Experiments} \label{S:experiments}
In this section, we describe the experimental procedure carried out for this paper. In order to quantitatively evaluate the performance of the presented breathing monitoring system, we deploy two nodes on opposite sides of a bed at a height of $0.85$ m, $1.95$ m apart from each other. The nodes are equipped with Texas Instruments CC2431 IEEE 802.15.4 PHY/MAC compliant transceivers \cite{CC2430}. The transceiver micro-controller units run a communication software and a modified version of the FreeRTOS micro-kernel operating system \cite{Freertos}, both developed by researchers at Aalto University. 

One of the nodes is programmed to transmit packets over each of the $16$ frequency channels defined by the IEEE 802.15.4 standard \cite{802_15_4} at the $2.4$ GHz ISM band. After each transmission, the node changes the frequency channel. The other node is programmed to receive the packets and to relay the data onward to a laptop for offline analysis. On average, the transmission interval between two consecutive packets is $1\text{ ms}$, with a standard deviation of $131$ microseconds between receptions. Thus, the sampling rate $T_s$ of each frequency channel is $16 \text{ ms}$ resulting in a sampling frequency $f_s = 62.5 \text{ Hz}$. The received packets are timestamped with a resolution of $1/32$ microseconds. 

During the test, a person is in the bed and breathing at a constant rate relying on a metronome to set the pace for exhalation and inhalation. The person is breathing at a rate of $12$ breaths per minute (bpm), i.e., $0.2 \text{ Hz}$. In each experiment, three different postures (lying on the back and on both sides) are tested to conduct breathing estimation in different poses. The changes in posture introduce motion interference to the measurements. In total, the test is repeated five times to verify reliability of breathing estimation using the methods presented. The experimental setup is shown in Fig. \ref{fig:experiment} and the experimental parameters are given in Table \ref{table:experiment_parameters}.

\section{results} \label{S:results}

In this section, we first give the conditional probability density fitting results for the HMM based motion interference detector and in Section~\ref{sec:estimation}, we evaluate the accuracy of breathing estimation using the experimental setup and methods proposed in this paper. In Section~\ref{sec:reduced_channels}, the effect of channel number to the accuracy of breathing estimation is empirically studied. Thereafter, we investigate the impact of the decimation factor $M$ and sampling frequency $f_s$ to the system performance. In Section~\ref{sec:posture} the effect of posture to breathing estimation is studied. We conclude the chapter in Section~\ref{sec:two_breathing} by experimentally showing that the breathing rate of two people can be estimated using the RSS measurements of a single TX-RX pair.

Throughout this section, the accuracy of breathing estimation is evaluated as the mean error of $\hat{f}$ in bpm,
\begin{equation} \label{eq:rmse}
     \varepsilon = 60\frac{1}{T}\sum_{t=1}^T (\hat{f}(t)-f),
\end{equation}
where $\hat{f}(t)$ is the $t^\text{th}$ estimate of the breathing frequency ${f}$, and $T$ is the total number of estimates. We do not consider the human induced errors in the experiments, i.e., phase noise is not further considered in the evaluations.  

\begin{table}
    \caption{HMM and conditional density parameters} % title of Table
        \centering % used for centering table
        \begin{tabular}{c c c} % centered columns (4 columns) \\
        \hline\hline\ %inserts double horizontal lines
        Parameter & Value & Description \\
        \hline  % inserts single horizontal line
        $P$ & $\begin{bmatrix}0.90 & 0.10 \\0.90& 0.10\end{bmatrix}$ & State transition probabilities \\ % inserting body of the table 
        $\pi_{s_0}$ & [1 0] & Initial probability  \\ % inserting body of the table
        $\sigma_1$ & 0.197 & Variance when in state $S_1$ \\ % inserting body of the table
	$\sigma_2$ & 2.385 & Variance when in state $S_2$ \\ % inserting body of the table
        \hline %inserts single line
        \end{tabular}
        \label{table:hmm} % is used to refer this table in the text
\end{table}

% ---------------------------------------------
% 	    	    HMM
% ---------------------------------------------
\subsection{HMM Conditional Density Fit}\label{sec-hmm-fit}

\begin{figure*}[t]
\begin{center}
\begin{tabular}{ccc}
\mbox
{
\subfloat[]{\includegraphics[width=\columnwidth]{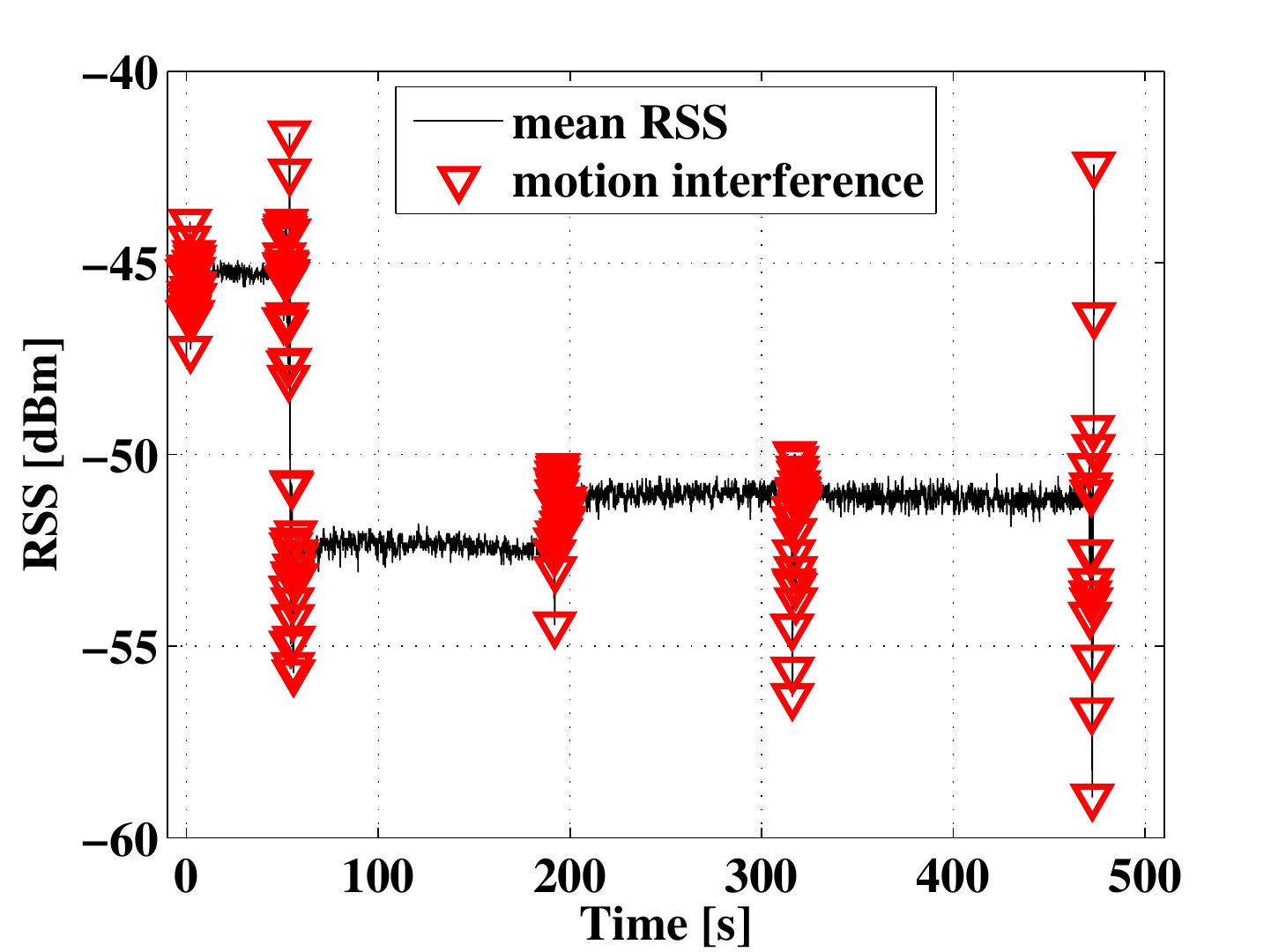}}
\subfloat[]{\includegraphics[width=\columnwidth]{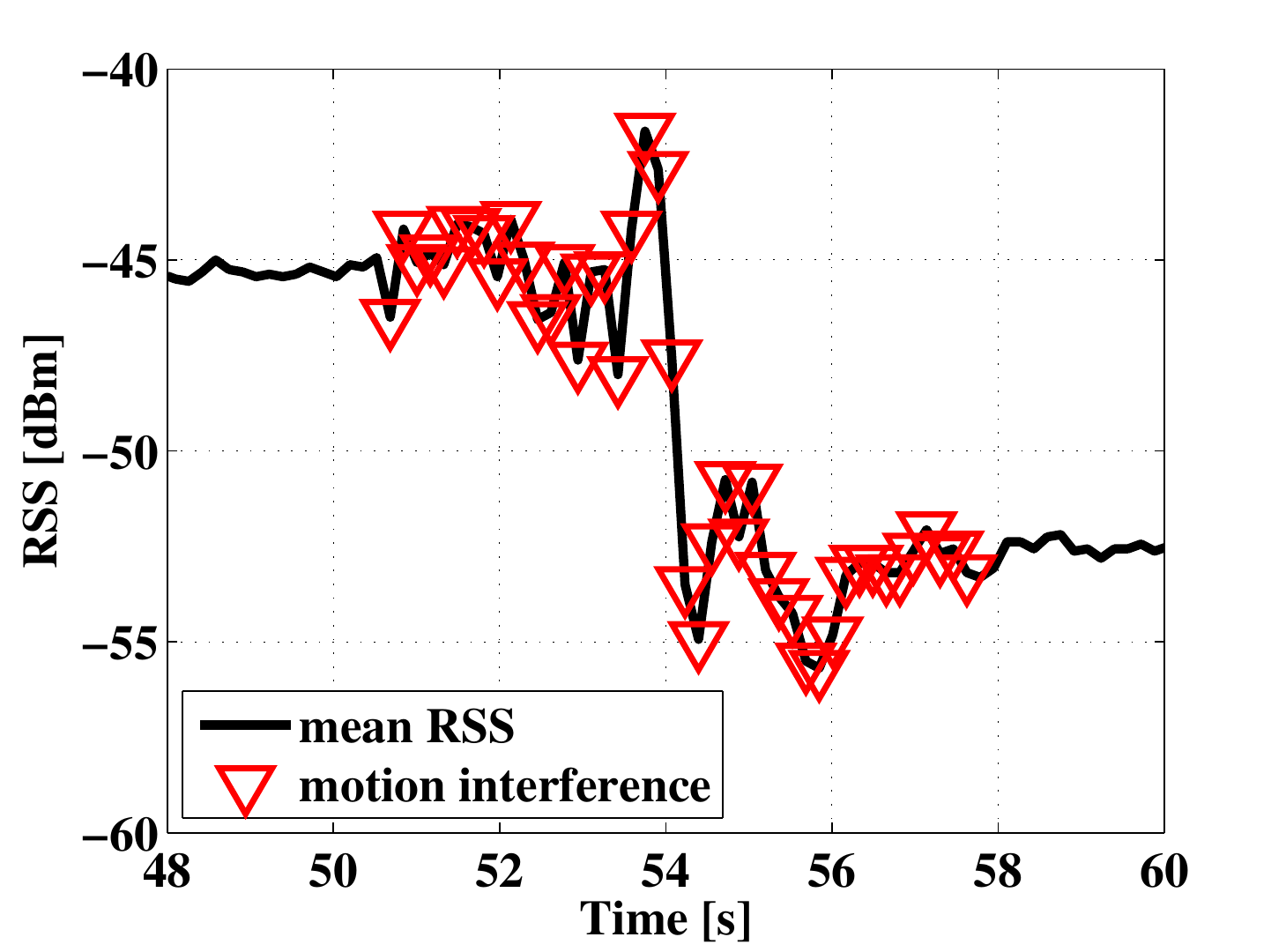}}
}
\end{tabular}
\caption{In (a), the motion interference instances (superimposed on the RSS measurements) triggered by the HMM in one breathing estimation experiment and in (b), a $12$ second time period from the test when the person gets into the bed and settles to lie on their back.} 
\label{fig:motion_interference}
\end{center}
\end{figure*}

\begin{figure*}[ht]
\begin{center}
\begin{tabular}{ccc}
\mbox
{
\subfloat[]{\includegraphics[width=\columnwidth*2/3]{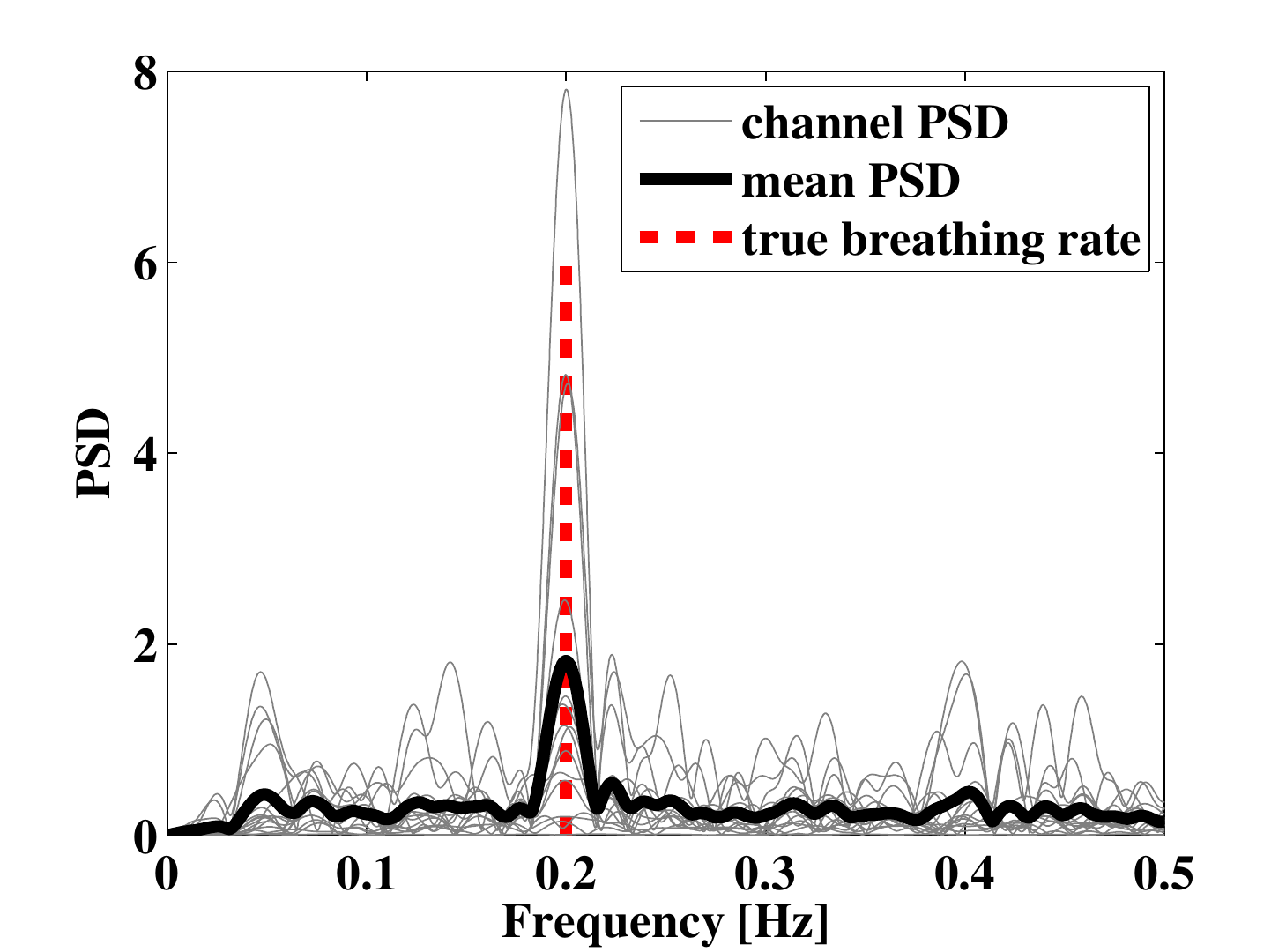}}
\subfloat[]{\includegraphics[width=\columnwidth*2/3]{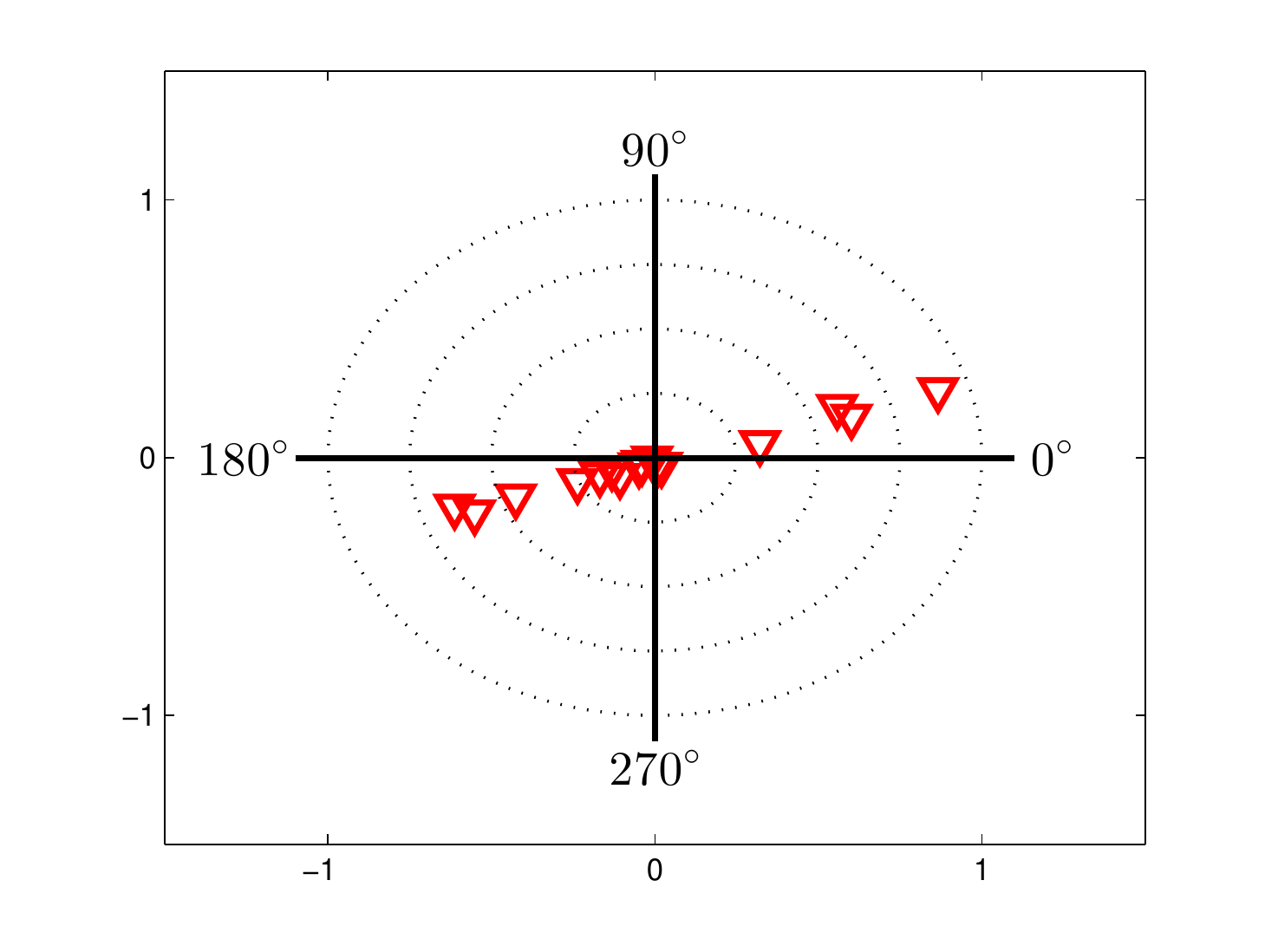}}
\subfloat[]{\includegraphics[width=\columnwidth*2/3]{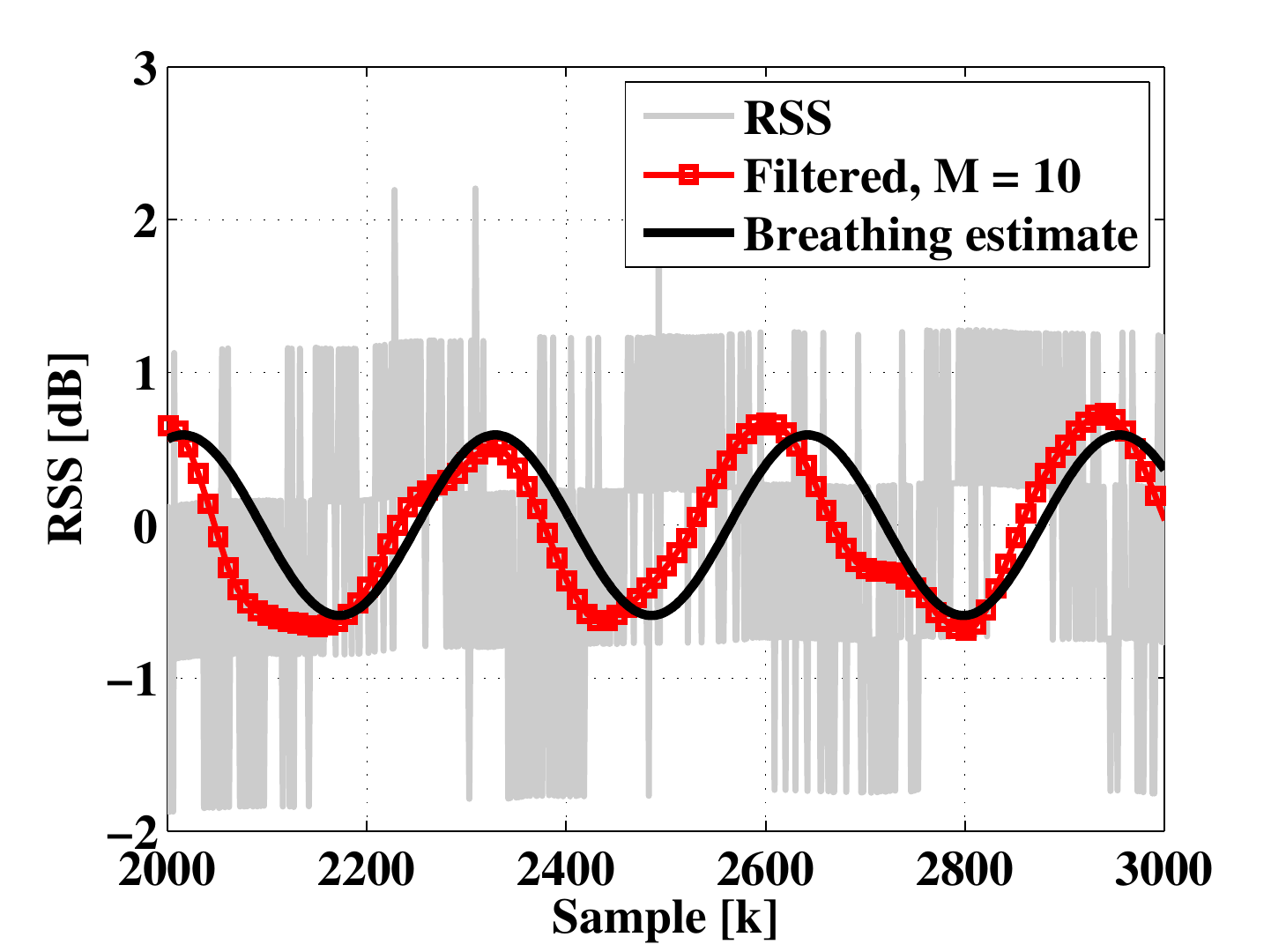}}
}
\end{tabular}
\caption{Calculated PSD in (a), amplitude and phase estimates on the $16$ different frequency channels in (b), and in (c), the measured and filtered RSS and the estimated breathing signal.} 
\label{fig:breathing_estimates}
\end{center}
\end{figure*}

We conduct two additional experiments in order to derive the conditional densities of the observations. In one of the tests, the person is breathing in between the transceivers but is otherwise stationary. In the other experiment, the person is continuously moving in between the nodes causing temporal fading. In both tests, approximately $18750$ RSS measurements on each frequency channel are collected. The data are used to determine the conditional densities of $x(k)$ in the two states. 

The empirical distribution of the observations in the absence of motion interference is shown in Fig.~\ref{fig:distributions} (a), and  the empirical distribution of the observations in the presence of motion interference is shown in Fig.~\ref{fig:distributions} (b). The empirical densities for both tests are obtained and tested against the theoretical distributions (given in Eq.~\eqref{eq:gaussian_distribution}) using the Kolmogorov-Smirnov test \cite{massey1969}. From both  densities, $1000$ samples are drawn and tested with respect to the theoretical distribution with a confidence level of $95\%$. For both tests, the hypothesis that the observations belong to the tested theoretical model is accepted. The p-values of the statistical tests are $38\%$ and $22\%$ for the Gaussian distributions when in state $S_1$ and $S_2$ in respective order.

Parameters of the HMM and the theoretical conditional densities are given in Table \ref{table:hmm}. The state transition probabilities of the HMM are derived from experimental data. It is to be noted that since the conditional densities reflect the observations of the two states accurately, the system is robust to changes in the state transition probabilities. In Fig. \ref{fig:motion_interference} (a), the motion interference instances triggered by the HMM in one of the experiments and in Fig. \ref{fig:motion_interference} (b), a $12$ second time period of the test are shown. The HMM identifies the motion interference instances reliably in all the experiments without causing false state transitions when the person is stationary but breathing.

%selected to reflect the reality. It is expected that most of the time the monitored person is stationary and therefore, we select $P_{1|1}=0.9$. On the other hand, motion interference is expected to be spurious and to last for short time intervals, thus we set $P_{2|2}=0.1$. 

\begin{figure*}[!thp]
\centering
\begin{tabular}{cc}
\subfloat[]{
     {\includegraphics[width=\columnwidth]{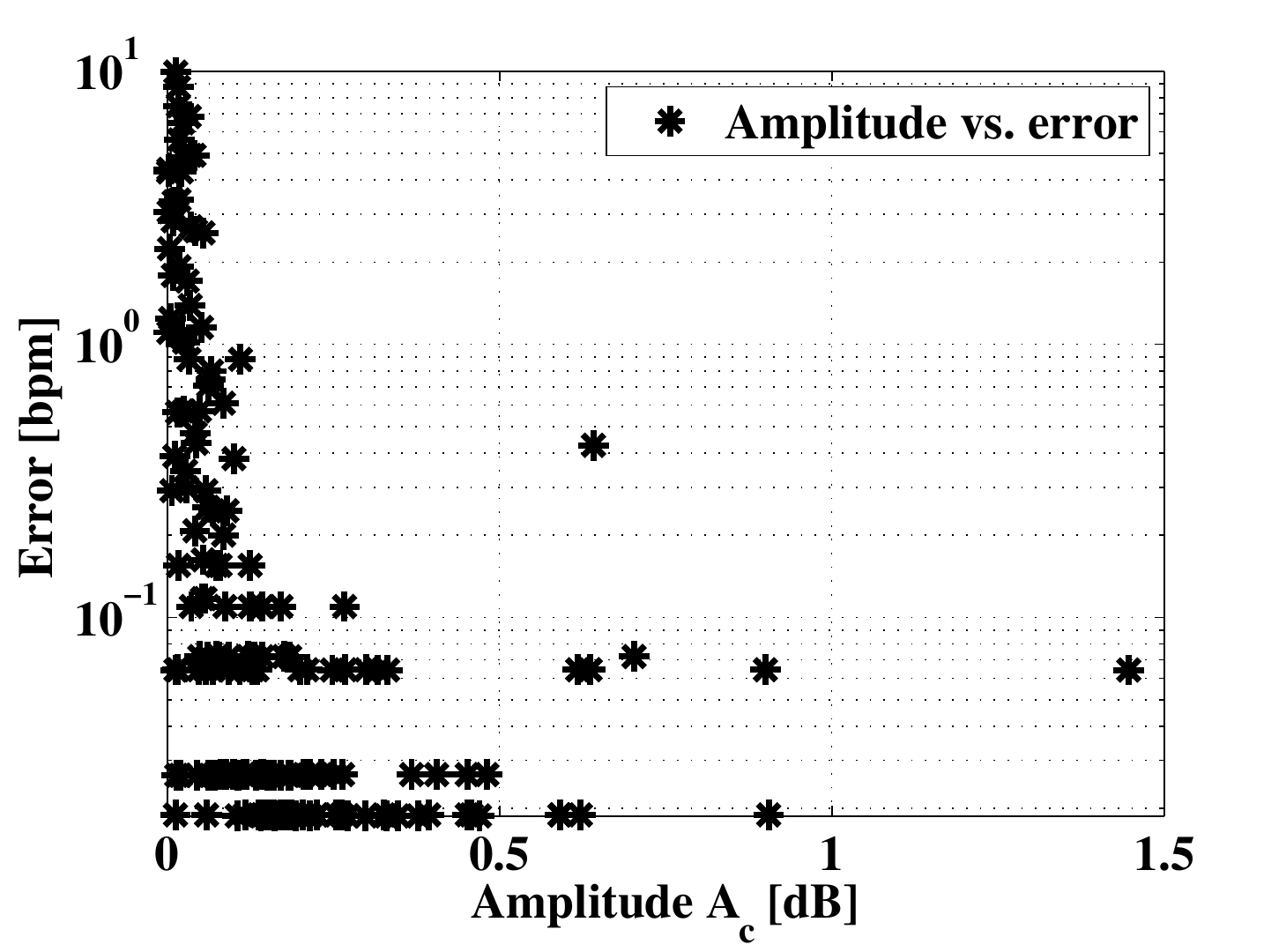}}
     \label{fig:amplitude_vs_error}}&
\subfloat[]{
     \includegraphics[width=\columnwidth]{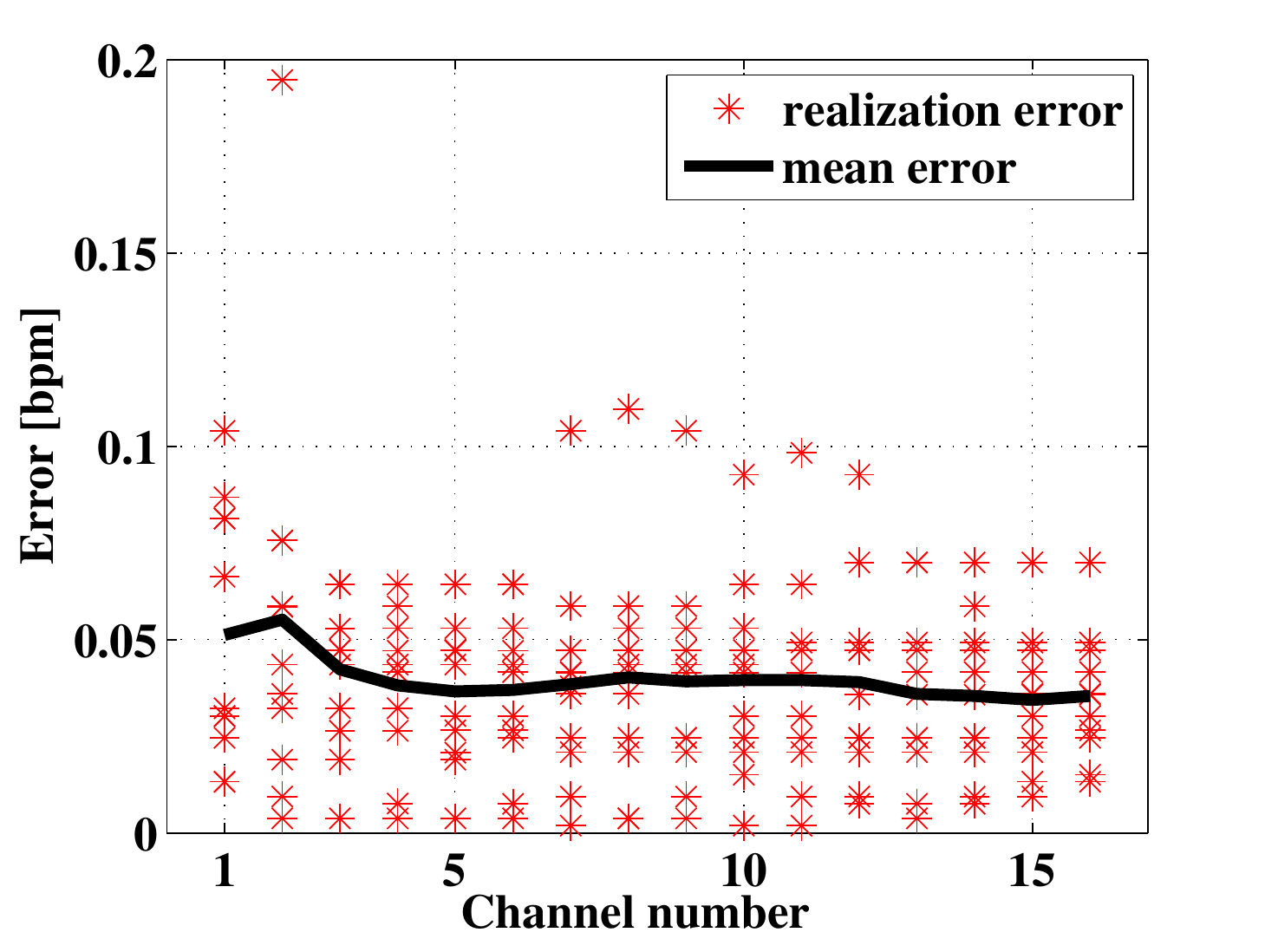}
     \label{fig:channel_errors2}}			
\end{tabular}
\caption{Respiration rate error as a function of estimated amplitude using a single channel shown in (a). In (b), breathing rate error as a function of channel number when using the highest amplitude channels to estimate the frequency.}
\label{fig:channel_ranking}
\end{figure*}

% ---------------------------------------------
% 	       ESTIMATING BREATHING
% ---------------------------------------------
\subsection{Estimating the Breathing Model Parameters}\label{sec:estimation}

The model parameters are evaluated using the latest $N$ measurements when in state $S_1$. One realization of breathing estimation is shown in Fig.~\ref{fig:breathing_estimates} (a), where the PSDs of the different channels are shown in gray and the average PSD of the channels is shown in black. The estimated breathing frequency of the time interval is $0.2005$ Hz, i.e., $12.03$ bpm an error of $0.25\%$ from the true respiration rate. Estimating the breathing rate is very accurate, and the mean error of the estimates is $0.03$ bpm with a standard deviation of $0.02$ bpm. For comparison, an end-tidal CO$_2$ meter, the gold standard breathing rate monitor used in hospitals \cite{cook2011major}, is accurate to $\pm 1$ bpm.

The polar plots of phase and amplitude estimates are shown in Fig.~\ref{fig:breathing_estimates} (b). In general, channels that are affected the most by breathing, measure a high amplitude. Thus, the amplitude of the model parameters can be used to detect the presence of a breathing person as proposed in \cite{Patwari2011}. In section \ref{sec:reduced_channels}, we exploit the amplitude estimates as a channel selection criterion and show that respiration rate monitoring is plausible using the measurements of a single TX-RX pair communicating on one frequency channel. 

It can be observed in Fig.~\ref{fig:breathing_estimates} (b) that the phase estimates are bimodal with two modes $180^\circ$ apart from each other. 
This observation is counterintuitive at first glance since the channel variation is caused by the same physical phenomenon, i.e., the breathing person in this case. However, exhaling may yield an increase or decrease in the RSS depending on the propagation channel and frequency channel of communication. Therefore, the RSS measurements of a particular channel might reach a maximum while another attains its minimum. We observed this behavior in all of the tests and in section \ref{sec:two_breathing}, we demonstrate that the phase estimates contain information about the number of people.

In Fig.~\ref{fig:breathing_estimates} (c), $\boldsymbol{\tilde{y}}(k)$, $\boldsymbol{r}(n)$, and $\boldsymbol{g}(n)$ are shown. The modeled signal $\boldsymbol{g}(n)$ deviates from the filtered signal $\boldsymbol{r}(n)$ due to phase noise. Even though we relied on a metronome to set a predefined pace for the respiration, due to the human-in-the-loop, it was not possible to assure that the person always exhaled and inhaled the same amount of air or that the duration of one breath cycle was always the same. In the long-run, these deviations average out and the estimated signal is sufficient for breathing monitoring.

% ---------------------------------------------
% 	       CHANNEL VS. AMPLITUDE
% ---------------------------------------------
\subsection{Channel Amplitude vs. Error}\label{sec:reduced_channels}

In this section, we investigate the relationship of the estimated signal amplitude and breathing rate error. First, we calculate the error for each of the three postures and five experiments. Then, we analyze the respiration rate errors separately for each of the $16$ frequency channels. The plot of breathing rate error as a function of estimated amplitude of the $240$ estimates is shown in Fig.~\ref{fig:channel_ranking}~(a). Clearly, the channels yielding higher amplitude estimates result in lower breathing rate errors. Channels that estimate an amplitude higher than $0.1$ dB, all result to an accuracy of one bpm or better. Respectively, if an accuracy of $0.1$ bpm is desired, on average, it can be achieved with channels that estimate an amplitude of $0.2$ dB or higher. Thus, the signal amplitude offers a good metric for channel selection which we investigate in the following.

We analyze the estimated amplitude as a channel selection criterion by removing channels yielding the lowest amplitude estimates. We increase the number of removed channels sequentially. For each channel number, we estimate the respiration rate and calculate the accuracy of breathing monitoring in each posture and experiment as shown in Fig.~\ref{fig:channel_ranking}~(b). In the figure, the breathing rate errors for each channel number and experiment are shown with stars, and the solid line represents the mean error. As can be seen, when the estimated signal amplitude is used as a channel selection criterion, the breathing can be accurately monitored even using the measurements of a single channel. Using only one frequency channel, the respiration rate can still be estimated with a mean error of $0.05$ bpm.

% ---------------------------------------------
% 	          SAMPLING
% ---------------------------------------------
\subsection{Sampling}\label{sec:sampling}

\begin{figure*}[t]
\begin{center}
\begin{tabular}{ccc}
\mbox
{
\subfloat[]{\includegraphics[width=\columnwidth*2/3]{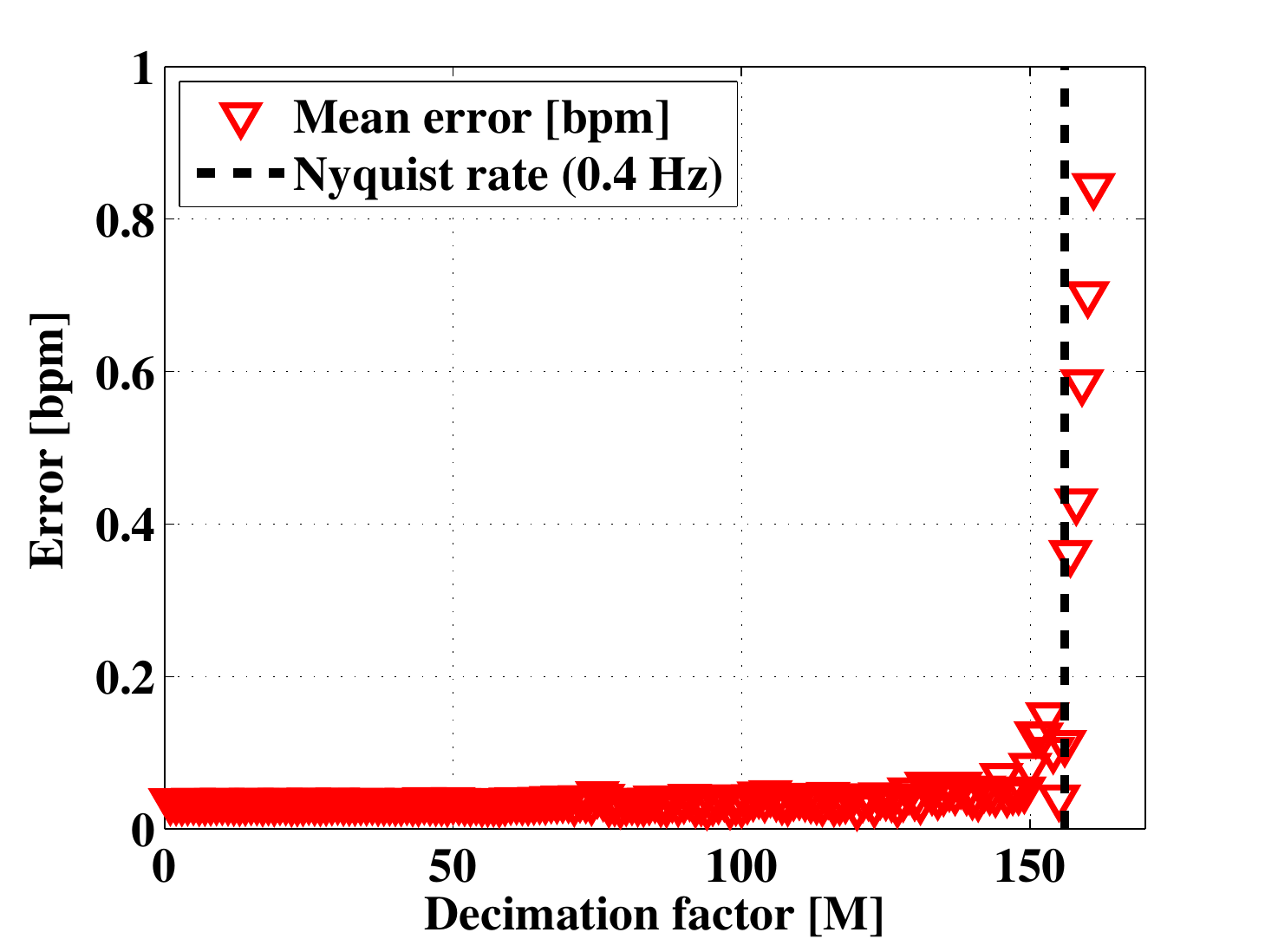}}
\subfloat[]{\includegraphics[width=\columnwidth*2/3]{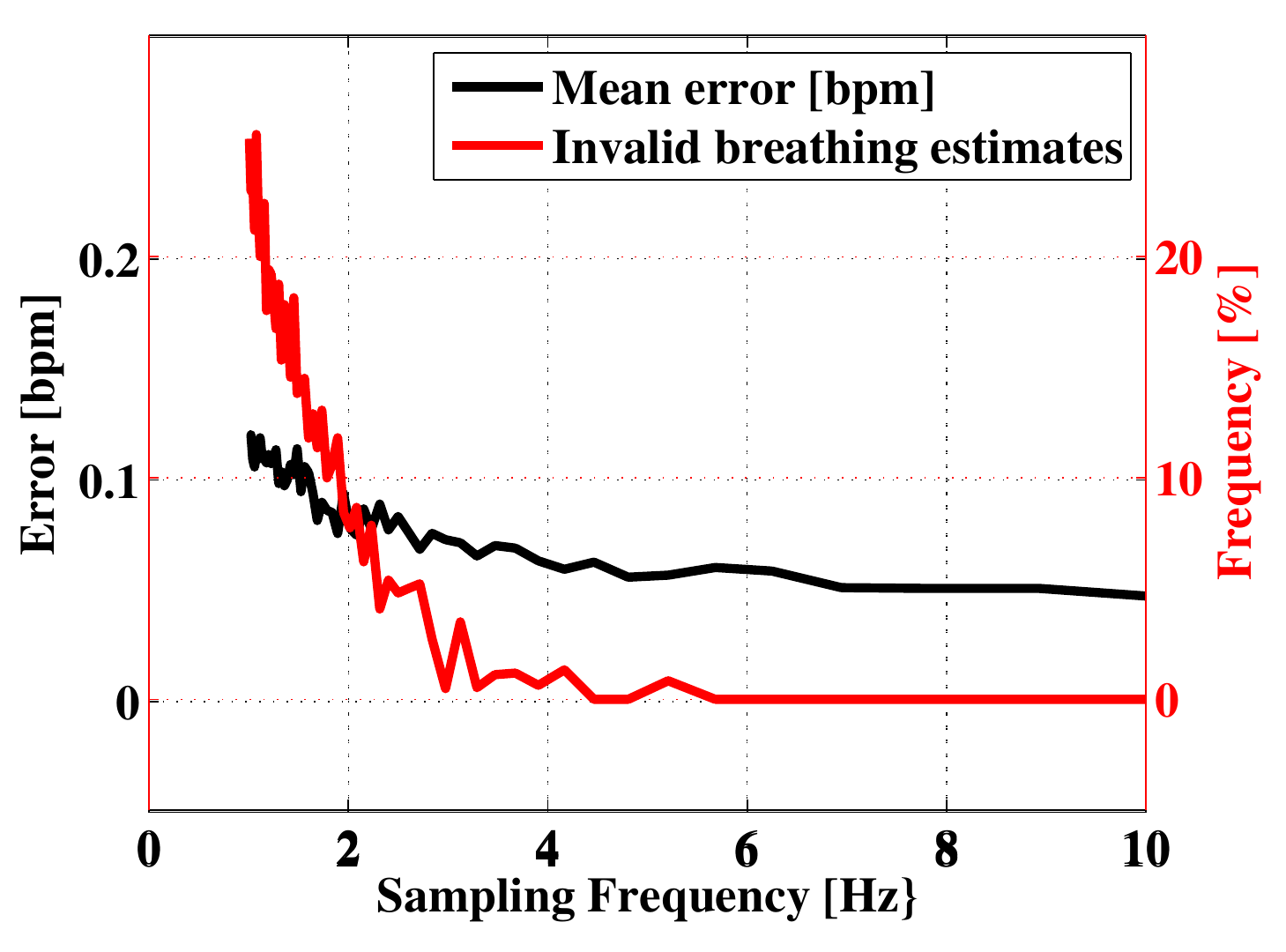}}
\subfloat[]{\includegraphics[width=\columnwidth*2/3]{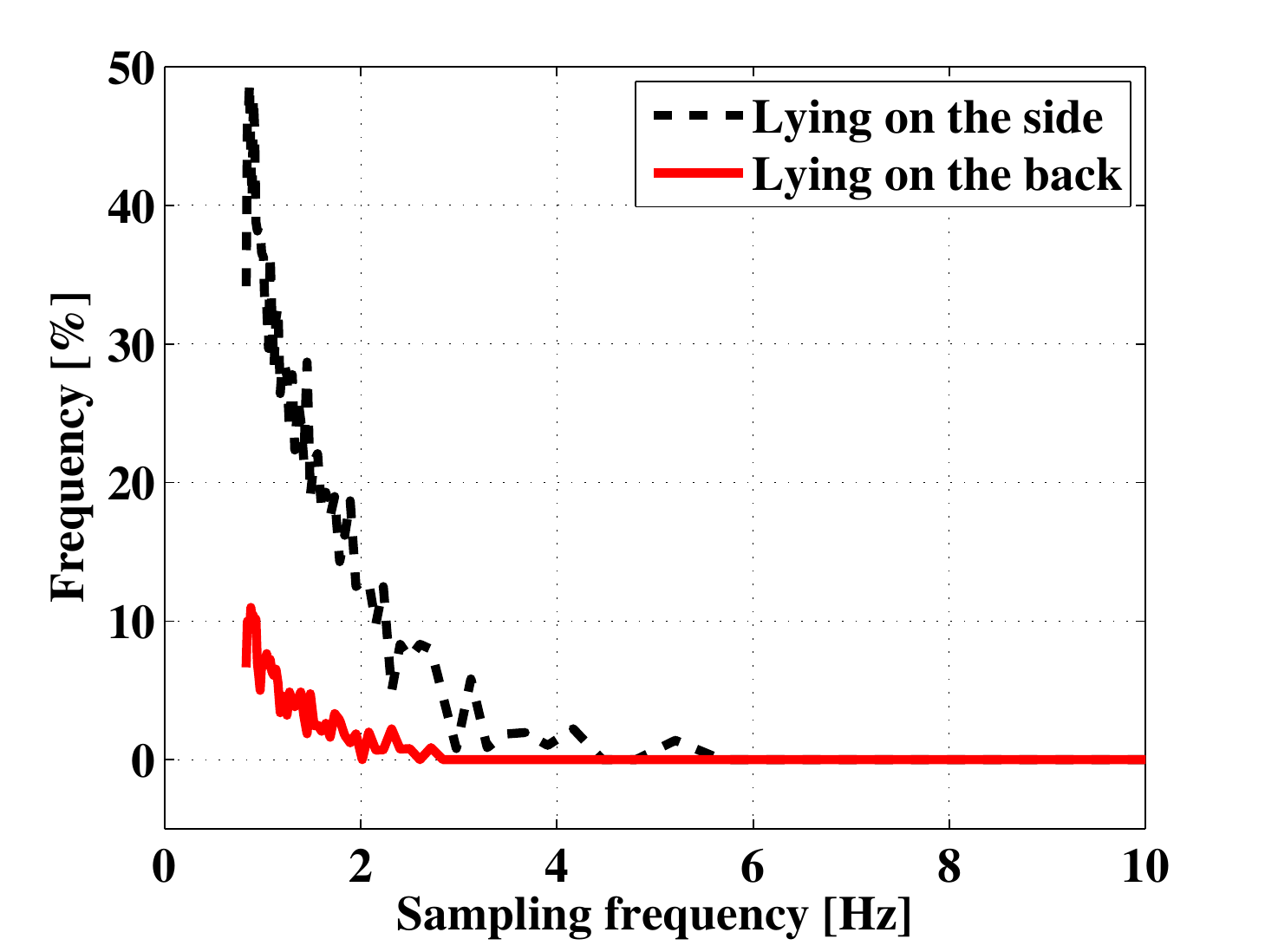}}
}
\end{tabular}
\caption{The effect of decimation factor $M$ presented in (a). In (b), respiration rate error and the percentage of invalid breathing estimates as a function of sampling frequency $f_s$. The percentage of invalid breathing estimates when the person is lying in different postures as a function of $f_s$ is illustrated in (c). } 
\label{fig:sampling}
\end{center}
\end{figure*}

In the following, we analyze the effect of decimation factor $M$ and sampling frequency $f_s$ to breathing estimation. The respiration rate error as a function of $M$ is shown in Fig.~\ref{fig:sampling}~(a). The decimation factor does not decrease the accuracy of breathing estimation considerably as long as the new sampling frequency $f_s/M$ is above the Nyquist frequency which in our case is $0.4$ Hz. However, the effect of the decimation factor to the computational overhead is significant, since not only is the number of measurements lower ($N/M$) but also the number of FFT points can be reduced to achieve the same resolution. For example, the average computation times of breathing estimation, using decimation factors of $M = [1, 10, 50, 100]$ are $[1.107, 0.081, 0.025, 0.024]$ seconds on a standard laptop computer, a reduction of $[0.0 \%, 92.7 \%, 97.7 \%, 97.9 \%]$ in computation times. Therefore, especially when decimation is exploited, implementing an online algorithm for breathing estimation is very possible.

In this paper, we exploit high sampling frequency to increase the resolution of the RSS measurements. However, it is not always possible to sample the channel with such a high rate e.g. when more sensors are used to add spatial diversity to the measurements. Therefore, it is important to investigate the effect of the sampling frequency to breathing estimation. In the analysis, we decrease the sampling rate of the original RSS measurements, i.e. 
$$\boldsymbol{\hat{y}}(l) = \boldsymbol{y}(\Delta k),$$where $\Delta=1,2, \cdots$. 

The results of breathing estimation using a lower sampling frequency along with the percentage of breathing estimates that yield $\varepsilon > 1.0$ bpm are shown in Fig.~\ref{fig:sampling}~(b). Reducing the sampling frequency beyond $6$ Hz has a negligible effect on breathing estimation. However, using sampling frequencies lower than $3$ Hz start to effect the results considerably as the breathing rate error increases the lower is the used sampling frequency. More notably, the number of failed breathing estimates increases rapidly with sampling frequencies lower than $3$ Hz. For example, already $25 \%$ of the estimates have an error higher than $1$ bpm while using a sampling frequency of $1$ Hz. Comparing the sampling frequencies and results with \cite{Patwari2011} ($f_s = 4.16$ Hz, $\epsilon = 0.3$ bpm) and \cite{Patwari2013} ($f_s = 2.34$ Hz, $\epsilon = 1.0$ bpm), we are confident that the achieved higher accuracy is mostly due to the advances enabled by the node and the used measurement setup. We exploit high sampling frequency, channel diversity, low-jitter periodic communication (standard deviation of $131$ microseconds between receptions), and accurate time stamping with a resolution of $1/32$ microseconds.

% ---------------------------------------------
% 		POSTURE
% ---------------------------------------------
\subsection{Effect of Posture}\label{sec:posture}

The RSS measurements are different for each posture and therefore, it is expected that the quality of breathing estimation varies with it. Interestingly, the pose has a negligible effect to breathing estimation when using a high sampling rate. However, the effect becomes evident when lower sampling rates are used as the resolution improvements due to over-sampling diminish. The percentage of breathing estimates that yield $\varepsilon > 1.0$ bpm error using lower sampling frequencies when the person is lying on their back and on their side is shown in Fig.~\ref{fig:sampling}~(c). 

It is noted that for our experimental setup breathing estimation is more robust when the person is lying on their back. In such a case, the largest chest movement is perpendicular to the line-of-sight. Therefore, the effect to the varying path length on the reflected waves is the largest. Correspondingly, when the person is lying on their side, the largest chest movement is parallel to the propagating RF signals. Thus, it has a significantly lower effect on the phase and amplitudes of the received multipath components resulting in lower $\boldsymbol{g}(k)$ amplitudes. Increasing the number of nodes and adding spatial diversity to the measurements could be exploited to increase the probability of successful breathing estimation for lower sampling frequencies.

% ---------------------------------------------
% 		TWO PEOPLE
% ---------------------------------------------
\subsection{Estimating Breathing of Two People}\label{sec:two_breathing}

\begin{figure*}[!thp]
\centering
\begin{tabular}{cc}
\subfloat[]{
     {\includegraphics[width=\columnwidth]{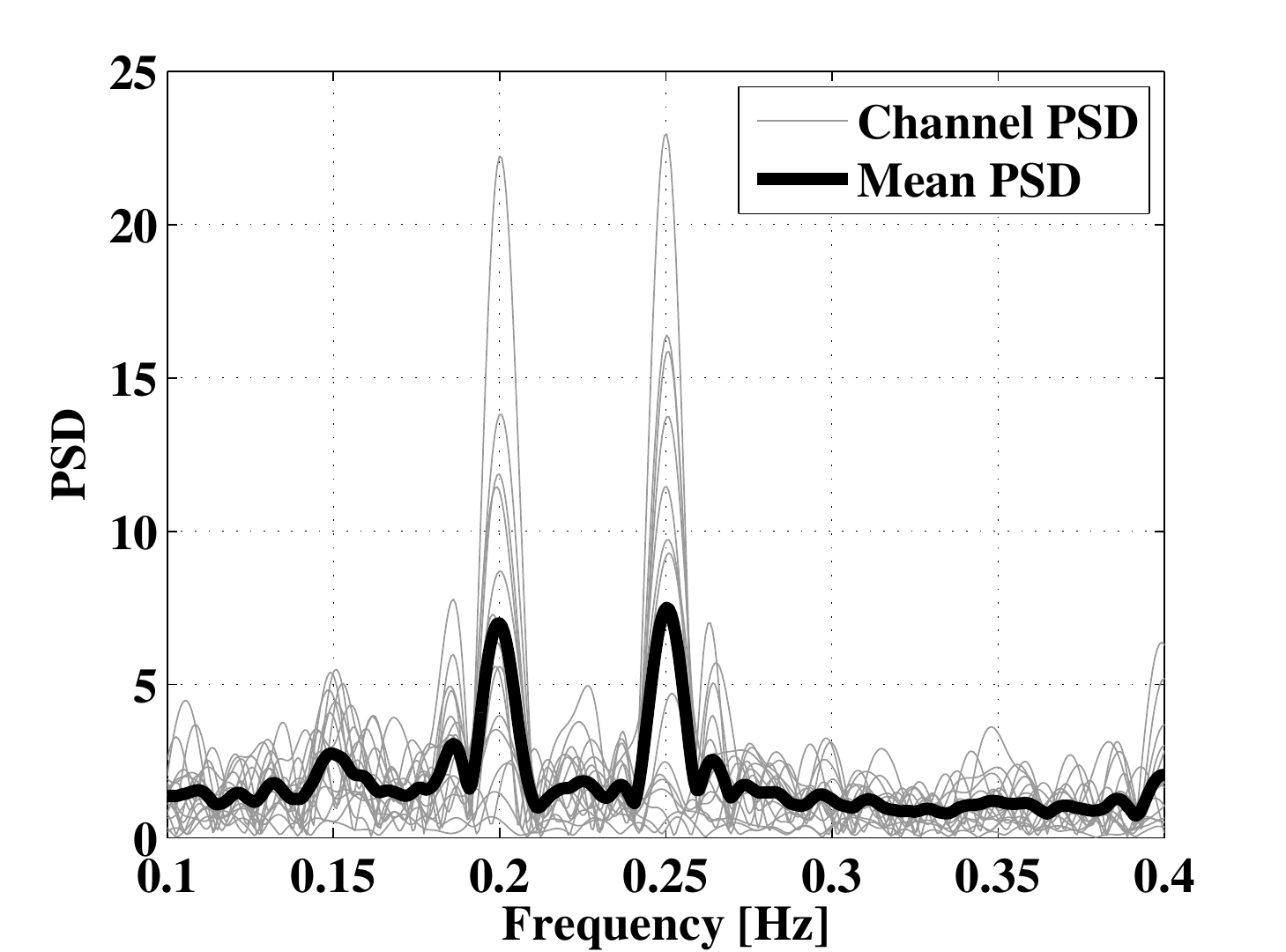}}
     \label{fig:psd_two_frequencies}}&
\subfloat[]{
     \includegraphics[width=\columnwidth]{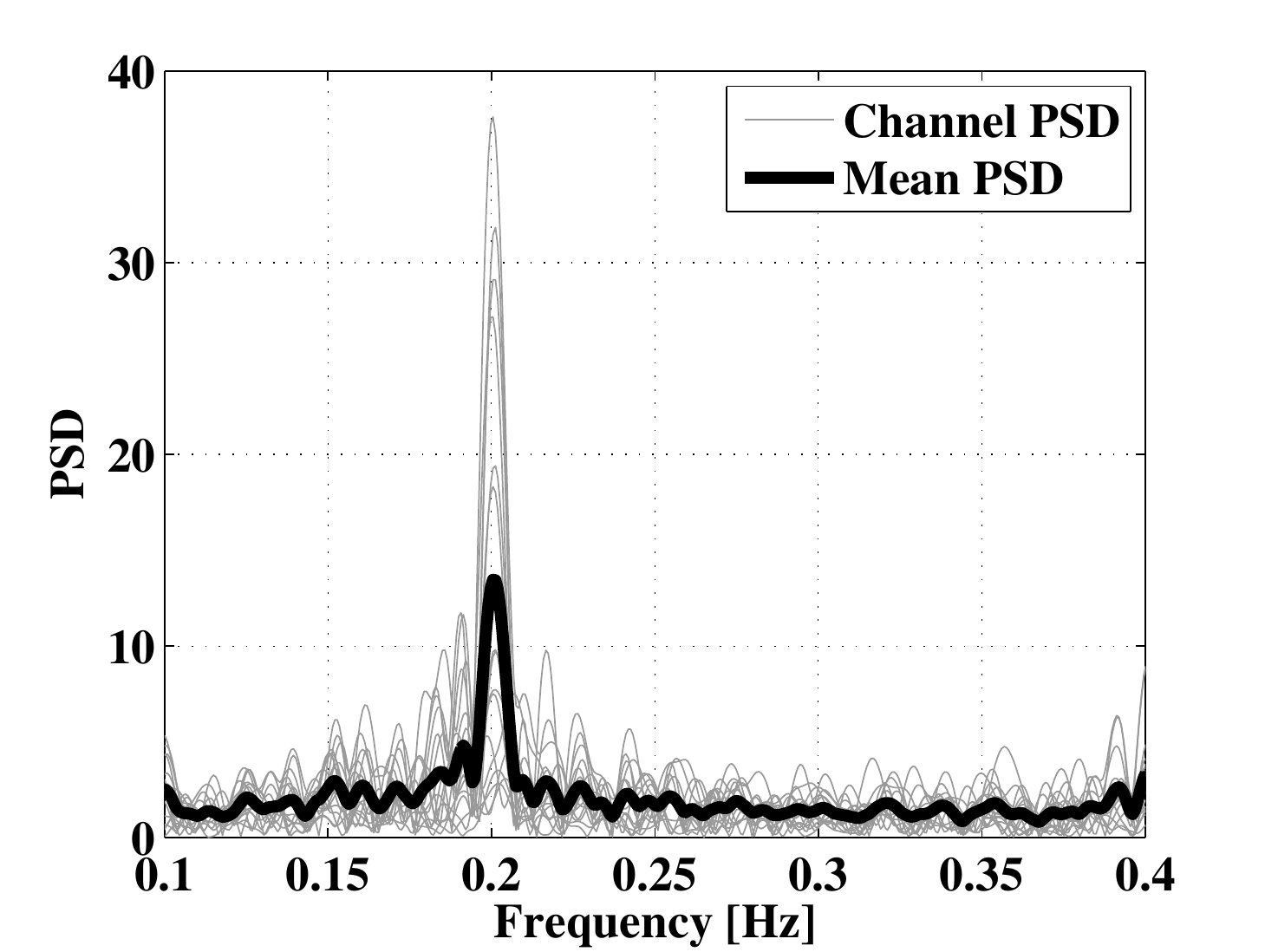}
     \label{fig:psd_two_phases}} \\
\subfloat[]{
     \includegraphics[width=\columnwidth]{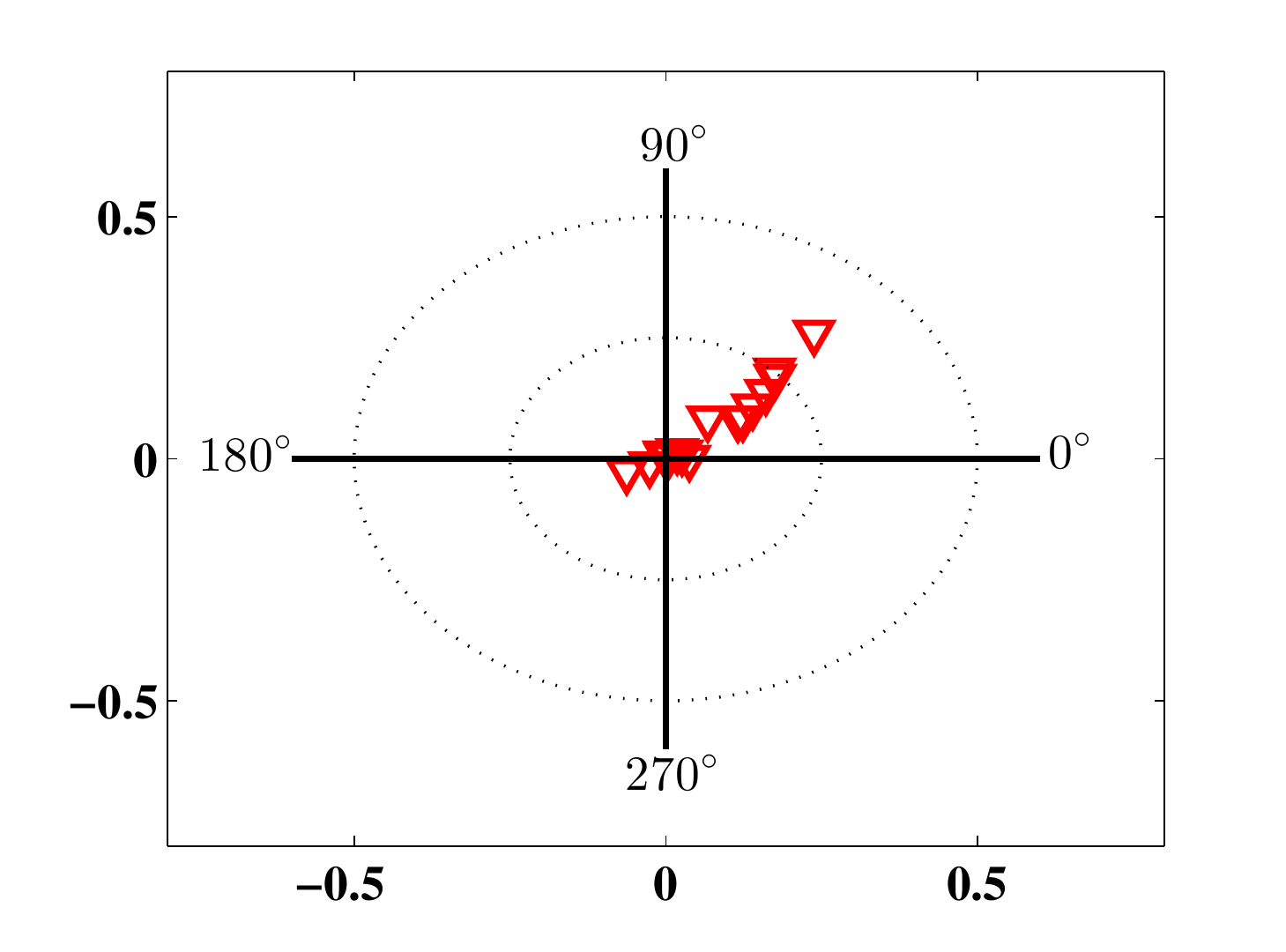}
     \label{fig:polarplot_two_frequencies}}&
\subfloat[]{
     \includegraphics[width=\columnwidth]{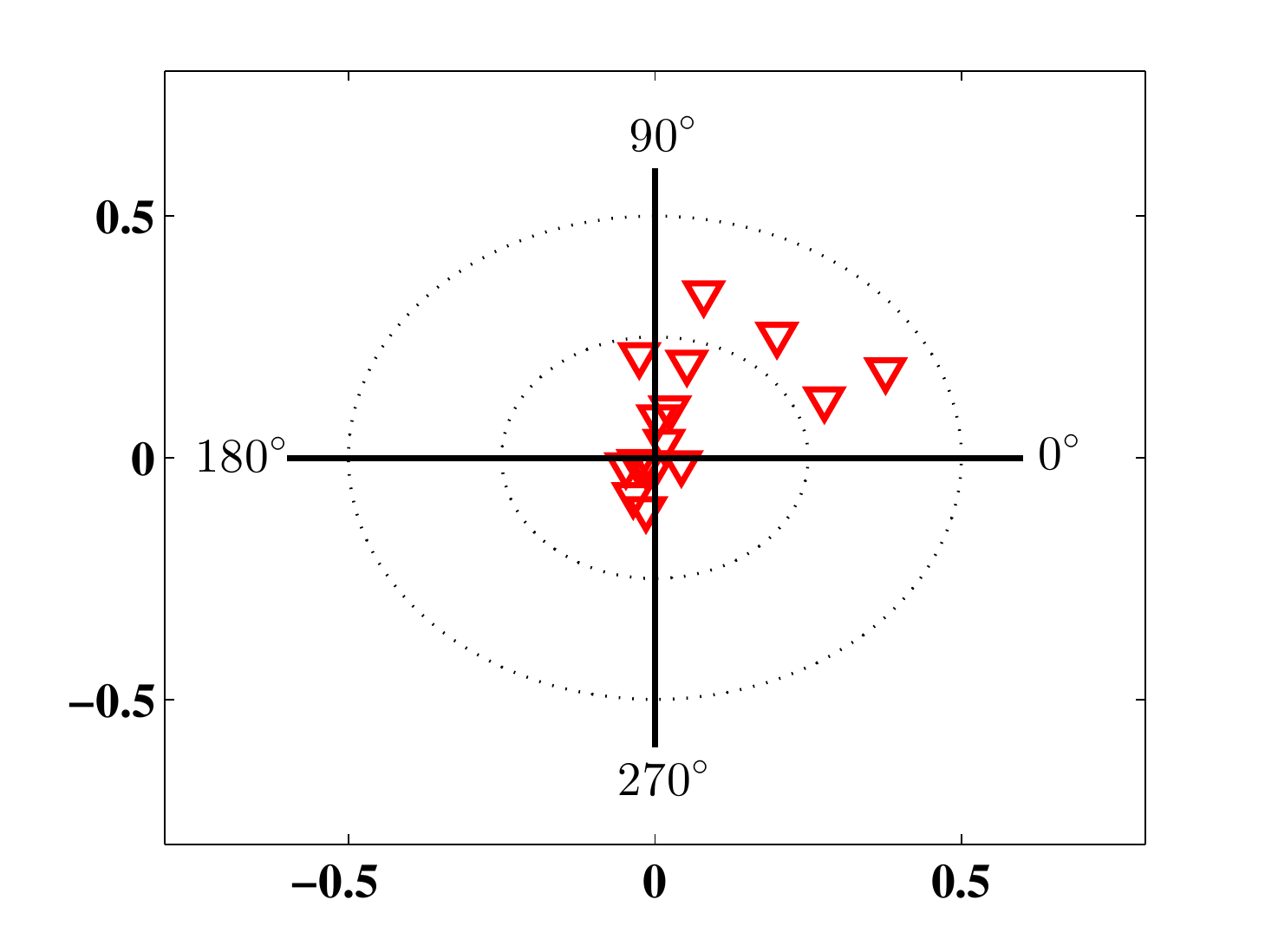}
     \label{fig:polarplot_two_phases}}
			
\end{tabular}
\caption{PSDs of two people breathing at different frequencies in (a) and at the same frequency in (b). Phase and amplitude estimates of two people breathing at different frequencies in (c) and at the same frequency in (d).}
\label{fig:two_people}
\end{figure*}

In the following, we demonstrate breathing monitoring of two people. However, we do not provide estimators for the model parameters when the signal is composed of multiples of sinusoids, each with unknown phase, amplitude, and frequency $-$ an important research topic that will be addressed in future work.

We conduct two experiments with the same setup as used for the single person test. In the first test, two people are breathing at different frequencies, i.e., $0.2$ and $0.25$ Hz which corresponds to breathing rates of $12$ and $15$ bpm. In the second test, the people are breathing at the same frequency ($0.2$ Hz) but at different phase. We time the breathing so to have a $90^\circ$ phase difference, i.e., when the other person has inhaled their lungs full, the other person has their lungs half empty. We want to know if the different frequencies and phases are resolvable from the RSS measurements.
 
The PSD, when breathing at different frequencies, is shown in Fig.~\ref{fig:two_people}~(a) and clearly there are two local maxima located at $0.1999$ Hz and $0.2503$ Hz, corresponding to an error of $0.006$ and $0.018$ bpm. As in the single person experiments, the phases of the different channels are bimodal as shown in Fig.~\ref{fig:two_people}~(c). In the experiment, the frequency separation between the breathing rates is high. Future work should investigate what is the required separation between the different breathing rates so that the local maxima are resolvable from the PSD.

When the people are breathing at the same frequency, as expected, the PSD only contains a single dominating frequency as shown in Fig.~\ref{fig:two_people}~(b). This maximum is at $0.2005$ Hz, an error of $0.03$ bpm compared to the true breathing rates. It is the phase now that contains the information of the two people, since the phase estimates are not bimodal anymore as shown in Fig.~\ref{fig:two_people}~(d). In the single person case, the phase estimates were always bimodal as shown in Fig.~\ref{fig:breathing_estimates}~(b) and therefore, the phase information can be used to identify that there may be more than one person in the monitored area. 

It is to be noted that in the experiments, only a single TX-RX pair was used to monitor the breathing. It is expected that increasing the number of nodes while communicating on multiple frequency channels would favor the problem of estimating the respiration rates of multiple targets.

% ---------------------------------------------
% 		CONCLUSIONS
% ---------------------------------------------
\section{Conclusions} \label{sec:conlcusions}

In this paper, we estimate breathing of a person using the RSS measurements of a single TX-RX pair and experimentally show that the breathing rate can be estimated with high accuracy.  We exploit channel diversity, low-jitter periodic communication, and oversampling to enhance the breathing estimates, but also make use of a decimation filter to decrease the computational requirements of breathing monitoring. In addition, we propose to use a hidden Markov model to identify the time instances when breathing estimation is not possible.

The results indicate that the breathing rate of a single person can be estimated with an accuracy of $0.03$ bpm, regardless of the person's posture. Further, we experimentally show that the amplitude of the estimated signals can be used to identify the channels that capture the breathing most reliably and propose to use it as a criterion for channel selection. Without influencing the accuracy considerably, we demonstrate that breathing estimation is possible using only the measurements of a single channel conditioned on all channels being monitored. In addition, we investigate the effect of sampling frequency and posture to the accuracy of respiration rate monitoring.

To the best of our knowledge, we are the first to show that breathing of two people can be monitored simultaneously. First, we show that different respiration rates leak to the RSS measurements and therefore, the breathing frequencies can be resolved from the PSD. Second, we investigate the situation where the people are breathing at the same rate but at different phase. The results show a clear dispersion in the phase estimates of the model parameters. This information can be used to identify that there are multiple people breathing in the monitored area.

We present an inexpensive alternative to non-invasive vital sign monitoring which can create new opportunities not only for improving patient monitoring in hospitals but also in home healthcare. Other opportunities of the proposed methods include: enhancing the life quality of elderly in ambient assisted living applications, to add context-awareness in smart homes, and in search and rescue for earthquake and fire victims.

%==================================================================================================%
%  Bibliography
%==================================================================================================%